\documentstyle[psfig,12pt]{article}
\begin{document}
\addtolength{\baselineskip}{2mm}
\newcommand{\refb}[1]{(\ref{#1})}

\title{A comparison of two operational wave assimilation methods}

\author{A.C. Voorrips\thanks{Supported by the Technology Foundation (STW).
Partly affiliated to Delft University of Technology, Faculty of Technical
Mathematics and Informatics, P.O. Box 5031, Delft, The Netherlands. E-mail:
voorrips@knmi.nl} \\
{\small Royal Netherlands Meteorological Institute (KNMI)} \\
{\small P.O. Box 201, 3730 AE De Bilt, The Netherlands} 
\vspace{5mm} \\
C. de Valk 
\thanks{Presently at ARGOSS, P.O. Box 61, 8325 ZH Vollenhove, The
Netherlands. E-mail: valk@argoss.nl} \\
{\small Delft Hydraulics } \\
{\small P.O. Box 177, 2600 MH Delft, The Netherlands } 
}

\date{\small{ Submitted to {\it The Global Atmosphere and Ocean System}}}

\maketitle

\newpage

\begin{abstract} \addtolength{\baselineskip}{1mm}

A comparison is carried out between two operational wave
forecasting/assimilation models for the North Sea, with the emphasis on the
assimilation schemes. One model is the WAM model, in combination with an
optimal interpolation method (OIP). The other model, DASWAM, consists of the
third generation wave model PHIDIAS in combination with an approximate
implementation of the adjoint method.

In an experiment over the period February 19 - March 30, 1993, the models
are driven by the same wind field (HIRLAM analysis winds), and the same
observation data set is assimilated. This set consists of a) spectra from
three pitch-and-roll buoys and b) Synthetic Aperture Radar (SAR) spectra
from the ERS-1 satellite. Three analysis/forecast runs are performed: one
without assimilation, one with assimilation of buoy measurements only, and
one with all data assimilated. For validation, observations from four buoys,
altimeter data from ERS-1 and Topex-Poseidon, and scatterometer data from
ERS-1 are used.

A detailed analysis of the "Wadden Storm" (February 20-22) shows the very
different nature of the two assimilation schemes: the wave and wind field
corrections of the WAM/OIP scheme are all in the vicinity of the
observations, whereas the DASWAM adjustments are more of a global nature.
The impact of some individual buoy and SAR observations is visualized. A
comparison of the performance of the two schemes is somewhat obscured by the
very different behaviour of the two first-guess runs.

A statistical analysis over the whole 39-day period gives the following
results. In a comparison with buoy observations it is shown that a positive
impact of wave data assimilation remains until about 12 hours in forecast
in both models. At the buoy locations, the impact of OIP assimilation in WAM
is larger, both at analysis time and in the short-term forecast. Comparison
with altimeter wave heights also shows a slightly larger impact of WAM/OIP
than of DASWAM. The impact of assimilation of buoy observations is larger
than the impact of satellite SAR observations, at least partly because of
the larger amount of buoy data.

The wind speed corrections applied by both assimilation schemes did not
significantly improve or deteriorate the quality of the winds, compared to
either platform or satellite wind measurements. For DASWAM, this is an
indication that a better representation of error covariances in the cost
function and a retuning of the wave model can further improve its
performance. 

\end{abstract}

\newpage

\section{Introduction}

Application of data assimilation to operational wave modelling is a quickly
developing subject. It was only a decade ago that first attempts were
reported to improve the wave forecast by correcting the wave field with
recent observations (Komen, 1985). Since then, the number of near-real time
available wave and wind observations has grown drastically because of the
launch of earth-observing satellites, notably the ERS-1 and ERS-2. At many
forecast centres, this has been the inspiration for the development of
methods which can make use of observations in the operational wave
forecasting cycle. In this paper, the performance of two of these methods
will be compared.

The first line of assimilation schemes consisted of sequential, time-
independent methods which were all specifically designed to assimilate
integral wave parameters, especially significant wave height (e.g. Thomas,
1988; Janssen et al, 1989; Lionello et al, 1995). These methods are
computationally cheap, and some success in improving the wave forecast has
been reported (e.g. G\"{u}nther et al, 1993). This has led to implementing
such a system in the operational wave analysis/forecast cycle at the
European Centre for Medium-Range Weather Forecasts (ECMWF). 

However, in other cases the impact of this type of systems has proved to be
small (Burgers et al, 1992; Mastenbroek et al, 1994; Bidlot et al, 1995).
As suggested by the last two studies, this may be caused partly by the fact
that significant wave height observations alone do not contain sufficient
information for a proper update of the wave spectrum. Recently, sequential
assimilation systems have been developed which are also capable of
assimilating observations of the full wave spectrum (Hasselmann et al,
1994b, 1996b; Voorrips et al, 1996; Breivik et al, 1996). Voorrips et al
showed the additional impact of the use of spectral information, by a
comparison with a method based on significant wave height assimilation.

Another line of algorithms consists of multi-time-level, variational
assimilation schemes, which minimise the misfit between model and
observations over a certain time interval (e.g., de Valk and Calkoen, 1989;
de Valk, 1994; Barzel and Long, 1994; Bauer et al, 1994; de las Heras et al,
1995; Holthuijsen et al, 1996; Hersbach, 1997). The advantage of
multi-time-level methods is that the model dynamics is taken into account
explicitly during the assimilation. The dynamics of wind sea and swell are
very different, the former being determined largely by the most recent wind
forcing, and the latter by wave propagation. This difference affects the
statistics of wind sea and swell: the spatial and temporal scales of
correlation for swell are variable and generally much larger than for wind
sea. Multi-time-level methods can take this explicitly into account by
letting the model completely re-generate the analysis after adjustment of
the forcing of the model. This will be an advantage in particular for the
analysis of swell, especially if assimilated observations and model output
locations of interest are spread over a large area, so the analyses can
benefit from remote observations. Another potential advantage of multi-time-
level methods is control over the way in which information from observations
made at different instants of time is integrated.

The drawback of these variational methods is that they are much more
demanding in terms of computer power than the single-time-level methods
described above. Therefore, full implementation of the variational technique
has so far only been applied to parameter estimation studies (Hersbach,
1997). Schemes which are used for state estimation still need
simplifications in order to restrict the computer time. De Valk and Calkoen
(1989) and Bauer et al (1994) apply a simplified wave model during the
minimization of the cost function. Holthuijsen et al (1996) avoid the
iteration in the minimization of the cost by assuming a parabolic dependence
of the cost function on the control parameters, and they apply a strongly
simplified approximation of the wind field in order to reduce the number of
control parameters.

In this paper, we compare the performance of two assimilation schemes, both
of which have been implemented in a wave analysis/forecasting cycle for the
North Sea. The first scheme, DASWAM, is a multi-time-level variational
method based on adjustment of the wind forcing history (de Valk and Calkoen,
1989; de Valk, 1994). The second scheme, OIP (Optimal Interpolation of
Partitions: Voorrips et al, 1996), is a sequential method which assimilates
wave spectra using the concept of {\it spectral partitioning} to reduce the
computational burden (see Gerling, 1992, and section \ref{oip}). The purpose
of the comparison is to assess whether the dynamically consistent analyses
of the variational scheme lead to more realistic analyzed fields and better
forecasts than the OI scheme, which uses fixed covariance fields. Attention
is paid not only to the wave fields, but also to the wind field update by
the assimilation schemes. Furthermore, the effect of different types of wave
measurements is studied.

For a 39-day period in February/March 1993, the wave models are run with the
same input winds, and the assimilation schemes are fed with the same wave
observations. The period includes the "Wadden Storm" (20-23 February), which
is a highly interesting event, in which significant wave heights up to 10~m
were measured in the North Sea. The observations come from directional wave
buoys and ERS-1 SAR. The results are validated against buoy measurements,
ERS-1 and Topex-Poseidon altimeter measurements, and ERS-1 scatterometer
data. 

In sections \ref{daswam} and \ref{oip}, the two assimilation/forecast models
are summarized. Section \ref{experiments} describes the comparison
experiment, of which the results are discussed in section \ref{results}.
Some conclusions are drawn in section \ref{conclusion}.

\section{Adjoint method (DASWAM)}
\label{daswam}

\subsection{Introduction}

The wave data-assimilation scheme DASWAM developed at Delft Hydraulics is
an example of a multi-time-level variational data-assimilation method. It
was implemented around 1990 for a wave model covering the North Sea and the
neighbouring part of the Atlantic Ocean (de Valk, 1994). Off-line tests were
reported in (Delft Hydraulics et al, 1994). In (Delft Hydraulics, 1995), a
real-time test producing analyses and forecasts every 6 hours over a period
of seven months was reported.

The basis of this approach is that possible errors in model input data are
corrected in order to minimize the misfit of the analyses to observations
and other available data, with the numerical model as a dynamic constraint
(Hasselmann et al, 1994a). The minimization is carried out by an iterative
descent method (Fletcher, 1987), using the model itself to evaluate the cost
associated to a trial value of the model inputs, and using the {\it adjoint
model} to compute the gradient of the cost to the model inputs, e.g.
(Luenberger, 1969). For numerical models with a finite-dimensional state
computed at discrete time levels, the adjoint model is nothing else than the
application of the chain rule to keep track of the gradient of the cost to
the state, also called {\it adjoint state} or costate. This section
discusses the wave model under consideration and specific choices made
concerning the input data to be corrected ({\it controls}), the {\it cost
functional} measuring the misfit of analyses to observations, and the
solution to the minimization problem.

\subsection{Model}

The wave prediction model applied by Delft Hydraulics is the third
generation model PHIDIAS implemented by Delft Hydraulics (van Vledder,
1994). It solves the evolution equation for the wave action density $A({\bf
k}, {\bf r},t)$:
\begin{equation}
\left\{
\frac{\partial}{\partial t}
+ ({\bf c_g} + {\bf U}) \cdot
  \frac{\partial}{\partial {\bf r}}
- ( \nabla_x \Omega ) \cdot 
  \frac{\partial}{\partial {\bf k}}
\right\}
A
= S_{in} + S_{ds} + S_{nl} + S_{bot}
\end{equation}
Here ${\bf k}$ the wave number, ${\bf r}$ position, $t$ time, ${\bf U}$
surface current, ${\bf c_g}$ group velocity and $\Omega$ angular frequency.
The right hand side terms $S_{in}$, $S_{ds}$, $S_{nl}$ and $S_{bot}$ are,
respectively, the source terms for wind input, dissipation through white-
capping, non-linear wave-wave interactions, and bottom dissipation. They are
taken equivalent to the original WAM source terms (WAMDI, 1988). The model
is thus quite similar to the original WAM model, the main difference being
that the action density is computed on a wavenumber/direction grid, whereas
in WAM, the variance density is computed on a frequency/direction grid. The
wavenumber grid consists of 25 grid points, with equal spacing of the
logarithm of the wavenumber, between 0.0036 rad/m and 1.01 rad/m,
corresponding to 0.03 Hz and 0.5 Hz in deep water. The directional grid
consists of 12 points with uniform spacing of 30$^\circ$.

For wave prediction for the North Sea region, the model was implemented on
a rotated spherical grid with pole in Venice; see figure
\ref{fig:modelregions}. This choice results in relatively high resolution
in the southern North Sea, and low resolution near Iceland. As a boundary
condition on the open boundaries of the model, for wavenumbers pointing to
the interior of the domain a vanishing gradient of the action density in the
direction normal to the boundary is imposed. In the absence of information
from other sources, this appears to be a reasonable choice, but substantial
errors can be expected in the region near the western boundary facing the
Atlantic Ocean. 

\subsection{Controls}
\label{controls}

In the application of variational inverse modelling within the wave
forecasting cycle to estimate the current state of the model, model inputs
to be corrected are naturally time-varying inputs (forcing, boundary
conditions, etc.). Forcing by wind is generally regarded as a major source
of error in numerical wave predictions and analyses. As a consequence, wind
fields over a preceding time-interval (the {\it assimilation window}) have
been chosen as controls. For the regional model for North Sea and north-east
Atlantic Ocean described earlier, the assimilation window was fixed at 72
hours. For this regional model, errors in boundary conditions are another
major source of error, but the correction of boundary conditions has not
been implemented so far.

The most obvious drawback of wind fields as controls is the huge dimension
of a wind field sequence. However, this presents only a minor problem in
practice (see below). Secondly, there is the problem of constructing
physically realistic wind fields. There are limits to the information about
wind fields obtainable from a restricted set of wave observations (de Valk,
1994), so we cannot expect that the wind field reconstructions will also be
consistent with the physics of the atmosphere. The most complete solution
is to couple the wave model to an atmospheric model and develop an
assimilation method for the compound model (Hasselmann et al, 1988),
provided that the difficult problem of selecting controls for an atmospheric
model has been sorted out. A provisional solution is to parameterize the
wind field sequence or to incorporate additional constraints on the wind
field sequence, such as geostrophy. In the current implementation of the
algorithm, smoothness of wind field corrections is imposed by a tensor
product B-spline representation of corrections $\Delta u$ and $\Delta v$ to
the wind vector components $u$ and $v$ (de Boor, 1987):
\begin{equation}
\Delta u (x,y) 
= \sum_{j,k} \alpha_u^{j,k} B^j(x) B^k(y)
\end{equation}
and
\begin{equation}
\Delta v (x,y) 
= \sum_{j,k} \alpha_v^{j,k} B^j(x) B^k(y)
\end{equation}
with $x$ and $y$ the longitudinal and latitudinal coordinates, respectively.
The spline coefficients $\alpha_u^{j,k}$ and $\alpha_v^{j,k}$ represent the
perturbation applied to the first-guess wind field and are zero at the start
of a data-assimilation run. For each 3-hourly wind field, a different set
of spline components is assumed. For each $j$, $B^j$ is a piecewise cubic
function with compact support determined by the knot points, which were
placed at every third mesh point of the wave model grid. The B-spline basis
functions are, except for shifts, identical in terms of the coordinates of
the rotated spherical grid. This results in smoother wind field adjustments
in the North. Other constraints are not imposed at present.

\subsection{Cost functional}

The cost functional should reflect the statistics of errors in observations
and in first guesses of model inputs. In general, a suitable choice is the
negative of the logarithm of the {\it posterior density} of the model
inputs, i.e. their conditional probability density relative to the
observations. When the cost function contains only terms expressing the
misfit to observations, minimizing it is equivalent to maximum likelihood
estimation of the model inputs. The misfit of wave analyses to observations
and the misfit of wind retrievals to first guess wind fields will appear in
separate terms of the cost functional. For most types of spectral wave
observations, appropriate cost function terms can be derived
straightforwardly by approximating the sea surface as stationary and
Gaussian. For directional wave buoys and measured SAR spectra, cost
functions have been implemented in this way. However, in the tests reported
in this paper, inverted SAR spectra were used for which no error statistics
are available (see section \ref{assimexp}). Also, the available parameters
of directional wave buoy spectra did not allow reconstruction of the 1st and
2nd order Fourier coefficients of the directional spreading functions which
are used in the cost functional term for these buoys (again, see section
\ref{assimexp}). Therefore, it was decided to formulate cost functionals for
both SAR and wave buoy data in terms of inverted directional wave spectra.
Directional spectra were derived from the wave buoy data of mean variance
density, mean direction and directional spread (Kuik et al, 1988) over 10
distinct frequency intervals by assuming a $\cos^{2s}$ directional
distribution for each of the intervals. Then these spectra were transformed
to the spectral grid of the model (but now in the frequency/direction
domain) using interpolation. The cost functional term for an inverted
spectrum from SAR and from a wave buoy used is a simple sum of squares
\begin{equation}
\sum_{i,j} 
\frac{\left[ F^{pred}(f_i,\theta_j) - F^{obs}(f_i,\theta_j) \right]^2 }
{\sigma^2}
\end{equation}
with $F^{pred}$ the model prediction of the wave variance spectrum at the
grid point and time-level of the observed wave variance spectrum $F^{obs}$.
The variance $\sigma^2$ was taken identical for all spectral bins for both
SAR and wave buoys. This simplistic approach has at least the advantage that
the effects of spectra from different instruments on the analysis is
independent of statistical assumptions, so they can easily be compared.
However, from a statistical point of view, it is not an optimal choice.

As mentioned in the previous section, deviations from first guess wind 
fields were represented by B-splines. Deviations from first guess wind 
fields were penalized by a cost functional term of the form of a sum of 
squares of all B-spline coefficients $\alpha_u^{j,k}$, $\alpha_v^{j,k}$ of
all wind fields, weighed by a  single constant. The weight was chosen
sufficiently small, so that the magnitudes of wind  field corrections did
not significantly restrict the influence of the wave  observations on wind
field reconstructions. This may not be very realistic,  but was motivated
by the situation that no adequate model of the covariance  structure of wind
prediction errors had yet been implemented. Aspects which  do affect the
outcomes are the spatial correlation length, determined by  the knot
distances of the B-spline basis, and the assumed mutual 
independence of the spline coefficients of consecutive 3-hourly wind 
fields. This latter assumption was found unrealistic in previous tests 
(Delft Hydraulics et al, 1994): time-series of wind field adjustments 
appeared rather bumpy, and storm systems were often found to be adjusted 
only at a single instant.

\subsection{Numerical solution}
\label{numerics}

To solve large-scale minimization problems, several efficient and
storage-effective descent methods are available which require only
functional evaluations (based on the forward model) and gradient evaluations
(based on the adjoint model). We implemented a limited-memory BFGS method
(Liu and Nocedal, 1988).

Still, applying this iteration scheme using a third-generation wave model
would require much more  computational effort and time than a forecast run,
because both the model and its adjoint need to be run over the assimilation
window (see \ref{controls}) at least once per iteration. At this stage,
approximation of the model within the minimization procedure seems the only
way out. The prerequisite of such an approximate model should be that for
wind fields close to the first-guess field, the response of the approximate
model should be close to the response of the real third-generation model.
In (Bauer et al, 1994), an approximation of the tangent linear model of WAM
is used. This is very good for wind fields very close to the first guess,
but deteriorates when the disturbance of the wind field becomes larger. We
chose instead to use a nonlinear second-generation wave model, tuned to the
third-generation model. This should give a better response when the true
wind field is rather far away from the first guess.

The wave field in the approximate second-generation model is discretized on
the same grid as PHIDIAS. Advection and wave dissipation are modelled in the
same way as in PHIDIAS. The main difference from PHIDIAS is the way in which
wave growth is modelled, which is the most expensive part of a third-
generation model. After advection, the procedure involves (a) extracting
wind-sea parameters from the advected spectra in a manner similar to
(Janssen et al, 1989); (b) updating these wind-sea parameters from friction
velocity, using growth curves tuned to the third-generation model; (c)
computing a parameterized wind-sea peak. In parallel, the advected spectrum
is dissipated as in WAM to deal with the swell components, and is then
combined with the computed wind sea peak by taking the maximum of both. 

Some details of the different steps in the computation follow below. In the
extraction of wind-sea parameters (a), the parameters extracted are the
total variance, mean wave period $T_m$ (defined in section \ref{evalmethod})
and mean direction $\theta_w$ over a wind-sea region of the spectral domain
determined from the friction velocity following (Janssen et al, 1989), with
friction velocity determined  from $U_{10}$ according to the Charnock
relationship. In (b), two growth curves are used: one for energy, and the
other for period, expressed as a function of  wave age (with all variables
made dimensionless using the friction velocity). The two growth curves were
fitted to calculations with a single-gridpoint re-implementation of WAM
without advection (De Valk  and Calkoen, 1989, Section 4.2). First an
effective wave growth duration is estimated from the wind-sea mean period
and energy and the friction velocity, defined as the average of the wave
growth durations computed from energy and from mean period. Then the energy
and mean period at the next time-level are computed each according to their
own growth curves. This procedure ensures that discontinuities in the growth
in the case of changing winds are minimized in both mean period and energy.
The parameterization of the wind-sea peak (c) is of the JONSWAP form
(Hasselmann et al, 1973; Komen et al, 1994, p. 187) but with different
coefficients to match the WAM spectra. A $\cos^2$ directional distribution
was assumed, independent of frequency. Weaknesses of the current version are
the directional relaxation and the behaviour of the spectral peak shortly
after a sudden decrease in wind speed. 

When comparing this second-generation model approximation with PHIDIAS or
WAM, differences in significant wave height in the order of 0.5-1 m were
frequently observed. Therefore, assimilation with this model may be
effective for correcting relatively large errors in wind fields but probably
not for correcting small errors, if it is assumed that the WAM model results
have a smaller error than the observed difference with the 2nd generation
approximation.

The second-generation approximation was not only used in the minimization 
of the cost function, but also to produce the analyses. The motivation is 
that due to differences between the second- and third-generation model, 
analyses produced using these models driven by the same wind fields may be
different. Using the second-generation model for the analysis has the
advantage that the wind field corrections have been obtained with the same
model.

As a result of the approximations made, producing the analysis over a 72 h 
interval performing 10 iterations to minimize the cost functional requires 
about twice the computational effort as running the forecast over a 36 h 
interval using the third-generation model. The overhead of data-assimilation
in comparison to a third-generation model forecast without data-assimilation
is therefore about 200~\%.

\section{Optimal interpolation of spectral partitions  (OIP)}
\label{oip}

\subsection{Introduction}

This section describes the wave data assimilation / forecasting system which
is currently at a semi-operational stage at KNMI (Voorrips et al, 1996;
Voorrips, 1997). Multi-time-level variational assimilation methods have been
used at KNMI for parameter estimation (de las Heras et al, 1995; Hersbach,
1997), but research for state estimation and wave forecasting is
concentrating on single-time-level, sequential methods. The latter
techniques are generally less computationally demanding and have been
applied successfully in meteorological forecast models. The drawback of such
techniques is, however, that the model dynamics cannot be incorporated
explicitly.

Previous experience has been obtained with the {\it optimal interpolation}
method applied to wave height and wave period measurements (Janssen et al,
1989; Lionello and Janssen, 1990; Burgers et al, 1992; Mastenbroek et al,
1994). Optimal interpolation is a statistical method which determines the
minimum error variance solution for the model state, by combining a model
first-guess field and observations, with prespecified forecast and
observation error covariances.

The method which is used in the present study is also an optimal
interpolation method, but an extended version which assimilates observations
of full wave {\it spectra}. The main spectral characteristics of the
spectrum are described by a reduced number of parameters, which limits the
computational burden which would be imposed by a full optimal interpolation
method. The method was devised and applied to ERS-1 SAR spectra by
Hasselmann et al (1994b, 1996b). Voorrips et al (1996) adapted this "Optimal
Interpolation of Partitions" (OIP) method for the use of pitch-and-roll buoy
data and applied it to the North Sea area. In their experiments, they showed
an improvement in wave analysis and forecast performance compared to the
scheme of Burgers et al (1992) in the North Sea case, due to the possibility
of assimilating spectral and directional details of the wave spectrum.

Since OIP is a sequential method, every assimilation step is performed only
for one fixed time level, processing all available observations near that
particular time. In a standard operational setting, observations are grouped
at 3-hourly intervals, so the assimilation scheme is called every three
hours. The computation time needed for the assimilation step is only a few
percent of that of a 3-hour run of the WAM model, so the additional cost of
assimilation with the OIP scheme is negligible.

The description of the WAM/OIP scheme below is concise. For an extensive
description we refer to Voorrips et al (1996).

\subsection{Wave model}

The wave model which is used is the WAM model, Cycle 4 (WAMDI, 1988;
G\"{u}nther et al, 1992; Komen et al, 1994). It solves the wave transport
equation
\begin{equation}
\frac{\partial F}{\partial t} +
\nabla \cdot ({\bf c_g} F) =
S_{in} + S_{ds} + S_{nl} + S_{bot}
\end{equation}
where $F(f,\theta,{\bf r},t)$ is the frequency-directional wave variance
density spectrum at location ${\bf r}$ and time $t$ and ${\bf c_g}(D({\bf
r}),f)$ is the group velocity depending on the local depth $D({\bf r})$ and
frequency $f$. The right-hand side represents the source terms due to wind
input, dissipation through white-capping, nonlinear wave-wave interactions
and bottom dissipation, respectively.

The model is implemented on a $\frac{1}{3}$ degree latitude $\times$
$\frac{1}{2}$ degree longitude grid (approximately 32 km grid spacing),
which includes the North Sea and part of the Norwegian Sea, see figure
\ref{fig:modelregions}. The Norwegian Sea is included mainly in order to
capture wave systems which are generated in this area, and which propagate
as swell into the North Sea afterwards. At each grid point, the wave
variance density spectrum is discretized in 25 frequencies ranging from 0.04
to 0.4 Hz, and in 12 directions. 

In the version of the model used in this study, the open boundary of the
model is not forced by externally generated wave fields. Instead, the
spectra at the boundary are defined to be equal to 90~\%  of the spectrum
at the neighbouring grid point within the model region, thus simulating a
wave growth with a finite fetch when the wind is blowing into the model
region. Currently, tests are being performed with open boundary forcing by
the global WAM version which runs at ECMWF.

\subsection{Assimilation method}

\subsubsection{Outline}

As mentioned above, computational efficiency of the OIP method is enhanced
by projecting the full wave spectrum onto a small number of parameters,
corresponding to so-called {\it partitions}, before the actual assimilation
is performed. This leads to the following step-wise approach:
\begin{itemize}
\item {\it Partitioning} of all observed and model spectra, i.e. division
of each spectrum into a few distinct segments. The physical interpretation
of each segment ("partition") is that it represents a wave system,
corresponding to a certain meteorological event (swell from a distant storm
in the past, wind sea which is generated by local wind). Every partition is
described by three mean parameters: its total energy, mean direction and
mean frequency.
\item {\it Cross-assignment} of partitions of different spectra: connect
partitions which are so close to each other in mean spectral parameters that
they can be supposed to represent the same wave system.
\item {\it Optimal interpolation} of the mean parameters from observed and
model partitions which are cross-assigned. Thus, an analyzed field of
partition parameters is obtained. 
\item {\it Update} of each spectrum locally, based on the first-guess
spectrum and on the analyzed partition parameters. 
\item {\it Update} of the driving wind field as well, if there is a wind
sea-component in the spectrum. 
\end{itemize}

\subsubsection{Spectral partitioning}

The concept of {\it spectral partitioning} was introduced by Gerling (1992).
It is a method to describe the essential features of a two-dimensional wave
spectrum $F(f,\theta)$ with only a few parameters, by separating the
spectrum into a small number of distinct segments, so-called partitions. The
partitioning is a purely formal procedure; however, the partitions can be
interpreted physically as representing independent wave systems. Details of
the formalism used to calculate the partitions can be found in Hasselmann
et al (1994b, 1996b) and in Voorrips et al (1996).

As said, the physical interpretation of a partition will be that of a wave
system, which has a different meteorological origin than other partitions
in the spectrum. The data assimilation scheme makes use of this
interpretation. The underlying assumption is that components of the
discretized spectrum which lie within a partition are fully correlated with
each other, whereas components from different partitions are uncorrelated.
This assumption is not entirely correct (wave systems may influence each
other through dissipation and non-linear interactions), but it is generally
a reasonable approximation. Having made this approximation, one can limit
oneself to calculating only a few integrated parameters of every partition,
and perform the assimilation on these integrated parameters rather than on
the full spectrum. The assimilation scheme as designed by Hasselmann et al
(1994b, 1996b) uses three parameters per partition:  the total energy of
each partition, the mean frequency and the mean direction.

Each partition is regarded to be either swell, or wind sea, or mixed wind
sea/swell according to a criterion involving the wind speed and the mean
wave vector of the partition (cf Voorrips et al, 1996).

\subsubsection{Partitioning of buoy spectra}

The original partitioning scheme (Hasselmann et al, 1994b, 1996b) can only
be applied to a full two-dimensional wave spectrum, such as a model spectrum
or an inverted SAR spectrum. Pitch-and-roll buoy data, however, contain only
the one-dimensional wave spectrum $E(f)$, plus limited information about the
directional distribution. To assimilate these data as well, an adapted
version of the partitioning scheme was developed (Voorrips et al, 1996)
which needs only $E(f)$ and the mean wave propagation direction
$\overline{\theta}(f)$ as a function of frequency. Tests with synthetic buoy
spectra which were extracted from full spectra showed good agreement between
the two partitioning schemes. 

\subsubsection{Cross-assignment of partitions}

The next step in the assimilation procedure is to merge the model
first-guess and observed partition parameters into an analyzed field of
parameters. We have assumed that different partitions within a spectrum are
uncorrelated, since they are created by different meteorological events. So,
we want to treat these partitions separately from each other in the
assimilation. On the other hand, partitions in different spectra (e.g.,
model and observed spectra, or two model spectra at different locations)
{\it are} correlated if they are created by the same event. Therefore, we
have to define a cross-assignment criterium between the partitions of two
different spectra, in order to decide whether a partition in one spectrum
represents the same wave system as a partition in the other spectrum. 

The criterium which is used is based on the distance in spectral space
between the mean parameters of two partitions. The ones which are closest
to each other  are cross-assigned. In case the number of partitions in the
observed and model spectra do not match, additional assumptions are needed.
For details, we refer to Voorrips et al (1996).

\subsubsection{Optimal interpolation of partition parameters}

When the cross-assignment is done, the mean parameters of the model and
observed partitions can be combined to obtain an analyzed field of partition
parameters. An important input for the OI procedure are the error
covariances of the errors in the observed and the model parameters. The
covariances were obtained by calculating long-term statistics of differences
betweeen observations and model forecasts. The observation errors were
assumed to be spatially independent. For the model forecasts, an error
correlation length of 200 km was found. The error variances of observations
and model values were taken to be equal and dependent on wave energy
(Voorrips et al, 1996).

\subsubsection{Update of wave spectra and wind field}

The analyzed partition parameters from the optimal interpolation are now
combined with the first-guess spectra to obtain analyzed spectra. Every
first-guess partition is multiplied by a scale factor and shifted in the
($f,\theta$) plane such that its mean parameters are equal to the parameters
obtained by the optimal interpolation. Small gaps in the spectrum which
arise by the different shifts for different partitions are filled by
two-dimensional parabolic interpolation.

When a wind sea partition is present in the spectrum, the driving wind field
is modified using a simple growth curve relation (Voorrips et al, 1996). The
new winds are used until the next wind field is read in, which is half a
wind time step later (90 minutes).

\section{Comparison of the DASWAM and OIP assimilation methods} 
\label{experiments}

\subsection{Assimilation experiment}
\label{assimexp}

The performance of the DASWAM and WAM/OIP forecasting/assimilation systems
(resp. sections \ref{daswam} and \ref{oip}) was compared by running the two
models for the same period, with the same input data (wind forcing and
assimilated observations), and by evaluating the results against independent
observations.

The selected period starts at 0 h GMT, February 19, 1993, and ends at 0 h
GMT, March 30, 1993. We chose this relatively long period (39 days) with
quickly changing meteorological conditions, as to be able to test the
methods in many different situations. The period includes the Wadden Storm
(February 20-23), in which wave heights up to 10~m were observed. To create
an initial condition at the start of the period, the models were run over
the previous two days without assimilation. Wind fields used were 3-hourly
analyses from the High Resolution Limited Area Model (HIRLAM; K{\aa}llberg,
1990), which is the operational weather prediction model at KNMI.

\begin{figure}[tp]
\centerline{\psfig{figure=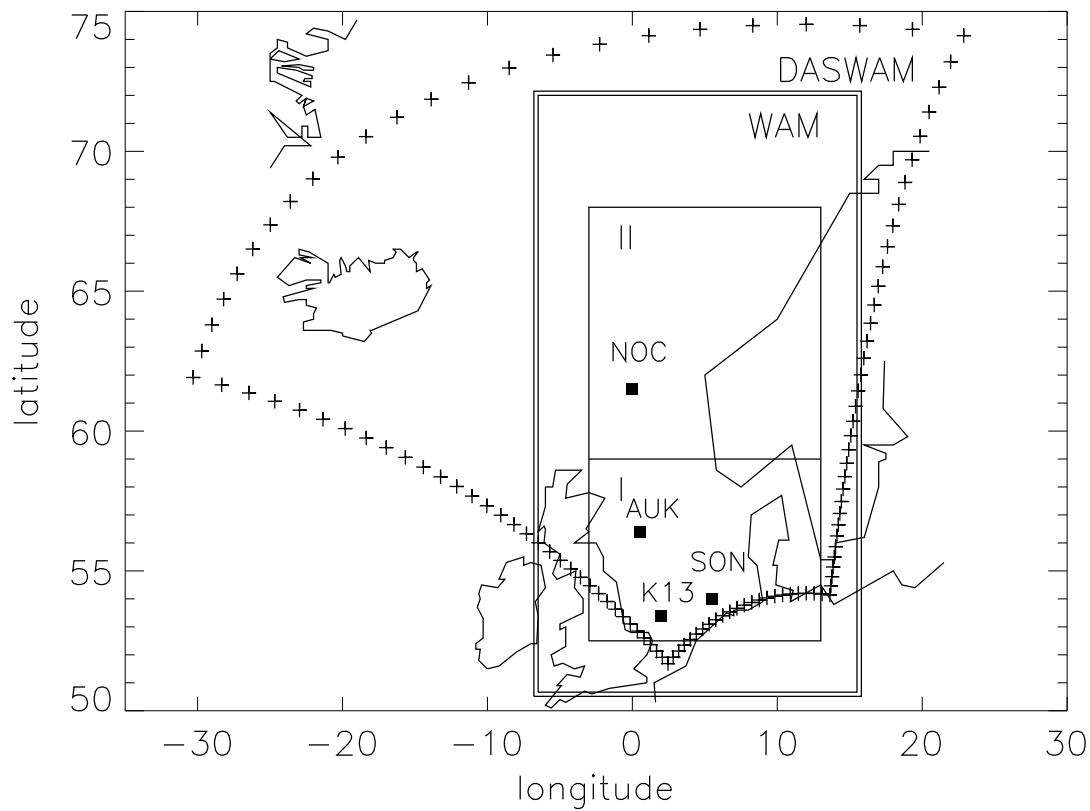,height=12cm}}
\caption{ \label{fig:modelregions}
Map of the area. Plusses indicate the DASWAM (Delft Hydraulics model) area.
Double solid lines are the boundary of the WAM (KNMI model) domain. Single
solid lines indicate areas I and II, which are used for the statistical
analysis of model results against altimeter observations. Filled squares
indicate the position of four wave buoys. NOC: North Cormorant; AUK: Auk
Alpha; SON: Schiermonnikoog Noord.}
\end{figure}

Two types of observations were used for the assimilation: 
\begin{itemize}
\item {wave buoy data: 3-hourly observations of spectral density, mean
direction and directional spread over 10 frequency bands between 0.03 and
0.5 Hz, from the locations North Cormorant (NOC), AUK and K13 in the North
Sea (see figure \ref{fig:modelregions}).}
\item {126 ERS-1 wave mode SAR spectra obtained from the Max-Planck Institut
f\"{u}r Meteorologie, Hamburg. The SAR spectra were inverted to wave spectra
using model output from a global WAM model as a first guess in the inversion
following Hasselmann and Hasselmann (1991) and Hasselmann et al (1996a).}
\end{itemize}

Once every 24 hours, starting at 00 h GMT, runs were performed which
consisted of two parts: (a) an "analysis" run over the previous time period,
in which data were assimilated; and (b) a 24 hours "forecast" run, starting
from the analyzed model state obtained after run (a). The time period used
for the analysis run (a) differs for the two assimilation methods. For the
OIP method, which is a sequential method, it was sufficient to perform a
24-hour analysis, starting from the model state obtained at the end of the
previous analysis run. For DASWAM, which is a multi-time-level variational
method, an analysis window of 72 hours was chosen, in order to assimilate
all relevant observations consistently with the model dynamics (see
subsection \ref{controls}). During the analysis run (a), DASWAM uses the
second generation approximate model (cf subsection \ref{numerics}), whereas
in the forecast run (b), it uses the third generation model PHIDIAS. WAM/OIP
uses the WAM model both in the analysis and in the forecast run (b). Note
that the "forecast" runs (b) were performed with {\it analyzed}, not {\it
forecast} wind fields. This was done to ensure that the only difference
between an analysis and a forecast lies in the assimilation of wave
observations; in this way, the impact of the assimilation is easily
monitored.

Three types of analysis runs (a) were performed, each with a subsequent
forecast run (b):
\begin{itemize}
\item A run without assimilation (NOASS run);
\item A run in which only the wave buoy data were assimilated (NOSAR run);
\item A run in which both wave buoy and SAR data were used (SAR run).
\end{itemize}
The NOASS run is used as a reference to study the impact of the real
assimilation runs NOSAR and SAR. The SAR run is done in order to study the
additional impact of the SAR data, added to the (much larger but less
homogeneous) amount of buoy data. Note that whereas the NOASS "assimilation
run" (a) and "forecast run" (b) are identical for the WAM/OIP system, they
are different for DASWAM: in (a), the second-generation wave model is used,
in (b) the third-generation PHIDIAS model is used. As will be seen in
section \ref{results}, the results of these can differ substantially.

Observations which were not assimilated are wave buoy spectra from the
location Schiermonnikoog Noord (SON) close to the Dutch coast (figure
\ref{fig:modelregions}), ERS-1 (OPR) and Topex-Poseidon (MGDR) radar
altimeter significant wave height and wind speed data, and ERS-1 (OPR)
scatterometer wind velocity data. These data were not not assimilated in
order to have a reference for the validation of the results.

\subsection{Evaluation method}
\label{evalmethod}

Results of the various assimilation runs were evaluated based on results for
the wave and wind parameters
\begin{itemize}
\item significant wave height $H_s$;
\item mean wave period $T_m = m_0/m_{-1}$ with $m_i = \int s(f,\theta) f^i
df d\theta$ the $i^{th}$-order spectral moment, or equivalently, the mean
wave frequency $f_m = 1/T_m$;
\item mean wave direction $\theta_w$;
\item low-frequency wave height, $H_{10} = 4 \sqrt{E_{10}}$, where $E_{10}$
is the total variance of all waves with periods above 10 s;
\item wind speed at 10 m above the surface $U_{10}$.
\end{itemize}
Model results of these parameters were validated against buoy measurements
at NOC, AUK, K13 and SON. The first three buoys were used during the
assimilation, so they are mainly of interest for forecast validation.
Additionally, significant wave height and wind speed corrections could be
checked against altimeter observations from the ERS-1 and Topex-Poseidon
satellites. Finally, the wind velocity corrections were compared to
measurements from the ERS-1 scatterometer. 

From ERS-1, the OPR products were used. A bias correction to the ERS-1
altimeter wave height data was applied using the formula (Queffeulou et al,
1994)
\begin{equation}
H_s^{corr} = 1.19 H_s^{OPR} + 0.19 [m].
\end{equation}
From TOPEX-POSEIDON, the GMDR data were used. The TOPEX wind speeds are also
known to have a bias compared to buoy observations (Gower, 1996; Komen et
al, 1996). Here we apply the correction of Komen et al:
\begin{equation}
\label{topexwinds}
U^{corr}_{10} = 0.877 U^{MGDR}_{10} + 0.39 [m/s].
\end{equation}

The evaluation will be split in two parts. In subsection \ref{wadden}, we
will study the most interesting part of the investigated period, namely the
Wadden storm (20-23 March). Through time series and maps, differences in the
characteristics of wave models and assimilation methods, and the impact of
SAR observations will be illustrated. In subsection \ref{statistics}, the
overall performance of the schemes will be presented through a statistical
analysis of model results against buoy and satellite data over the full
period.

\section{Results}
\label{results}

\subsection{The Wadden Storm, 20-23 February 1993}
\label{wadden}

\subsubsection{Wind fields}
\label{windfield}

\begin{figure}
\vspace{-2cm}
\centerline{\psfig{figure=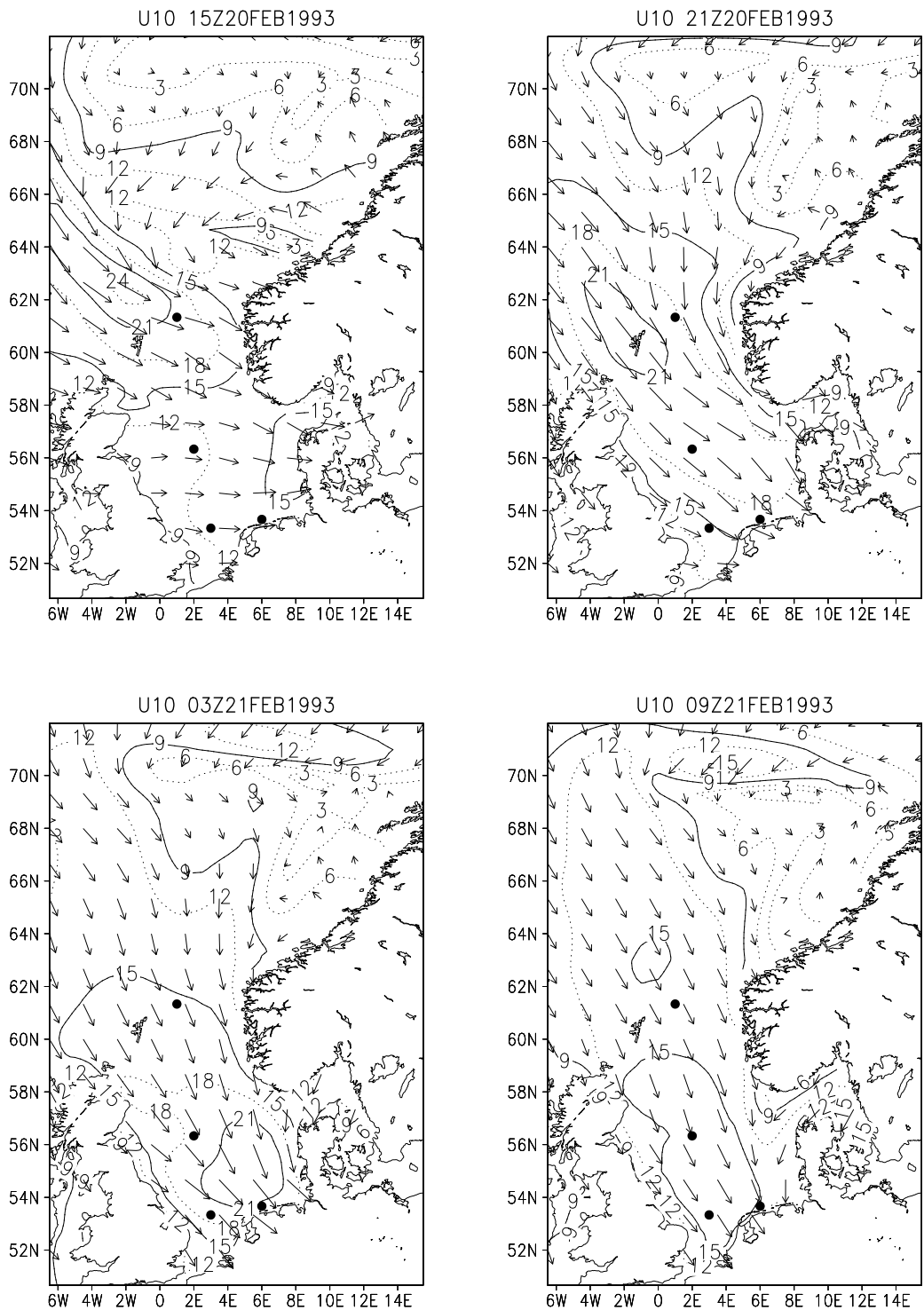,height=20cm}}
\caption{ \label{fig:winds_noda}
Wind fields at the maximum of the Wadden storm. Upper left: February 20, 15
h GMT. Upper right: February 20, 21 h GMT. Lower left: February 21, 3 h GMT.
Lower right: February 21, 9 h GMT.}
\end{figure}

\begin{figure}[t]
\centerline{\psfig{figure=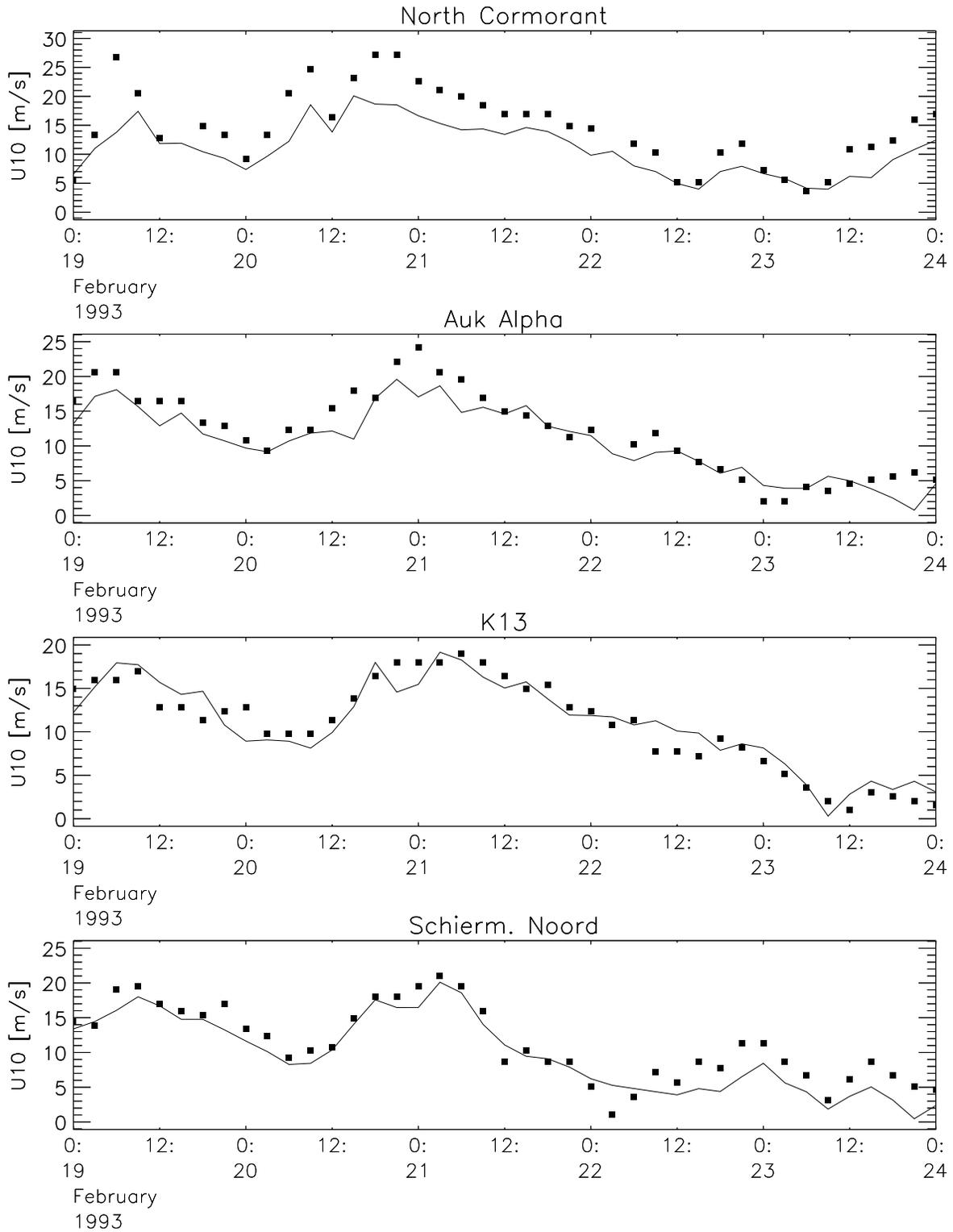,height=25cm}}
\vspace{-4cm}
\caption{ \label{fig:daswam_noda_1}
Modeled and observed wind speed at four locations for the period February
19-24, 1993 (the "Wadden Storm"). Markers: observations. Solid line: HIRLAM
model results.}
\end{figure}

Figure \ref{fig:winds_noda} shows four HIRLAM wind fields at the height of
the Wadden storm, from February 20, 15 h GMT until February 21, 9 h GMT.
Within 12 hours, the maximum of the north-westerly storm moves quickly from
the North of Scotland, through the North Sea to the German Bight. In figure
\ref{fig:daswam_noda_1}, measured and modeled wind speeds are given at four
North Sea locations, which are indicated in figure \ref{fig:modelregions}.
For the southerly locations K13 and SON, the agreement between model and
observations is satisfactory. At AUK, and especially at NOC, however, the
model wind speed is significantly lower than the platform observations. The
underestimation is almost 10 m/s at the maximum of the storm. 

Only few satellite wind speed measurements are available during the storm,
to complement the platform measurements. From these measurements, no
consistent bias of the model wind speed can be inferred, although
occasionally, large differences between model and observations are found.
An additional problem here is that the quality of altimeter measurements of
wind speeds above 20 m/s (which occur at the height of the storm) is not
well-known, due to lack of statistics in most comparisons between
conventional and satellite measurements (e.g., Gower, 1996; Komen et al,
1996). 

From the above, one can see two good reasons to use the Wadden Storm in a
study of wave data assimilation schemes. First, the extreme north-westerly
winds will generate high waves which can travel over a large distance into
the North Sea. This is a typical situation in which assimilation of data in
the northern and central North Sea can improve the wave forecast in the
South. Second, the poor quality of the model winds at the height of the
storm will leave sufficient room for the assimilation of wave observations
to have a large impact on the analysis and prediction of the sea state.

\subsubsection{NOASS runs}

In this paragraph, we compare the results of the various model runs without
assimilation. The difference between the three wave models used will be of
importance when comparing the performance of the data assimilation schemes,
in which we are mostly interested.

\begin{figure}[t]
\centerline{\psfig{figure=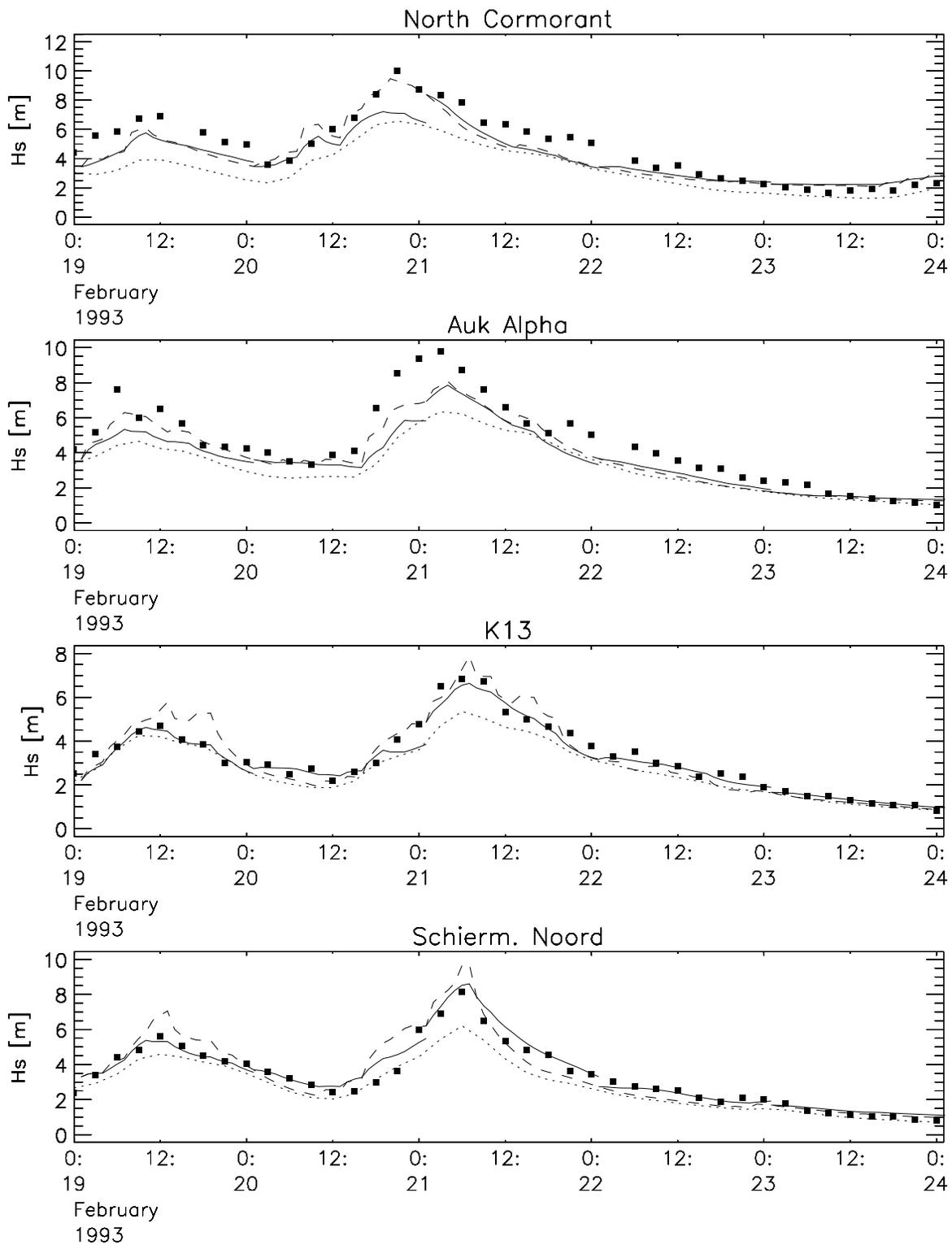,height=25cm}}
\vspace{-4cm}
\caption{ \label{fig:daswam_noda_2}
Modeled and observed significant wave height at four locations for the
period February 19-24, 1993. All model runs without assimilation. Markers:
observations. Dotted line: NOASS analysis run with WAM. Dashed line: NOASS
analysis run with DASWAM (2nd generation model). Solid lines: 24 hour NOASS
forecast runs with DASWAM (3d generation model), starting every day at 0 GMT
from the DASWAM NOASS analysis.}
\end{figure}

\begin{figure}[t]
\centerline{\psfig{figure=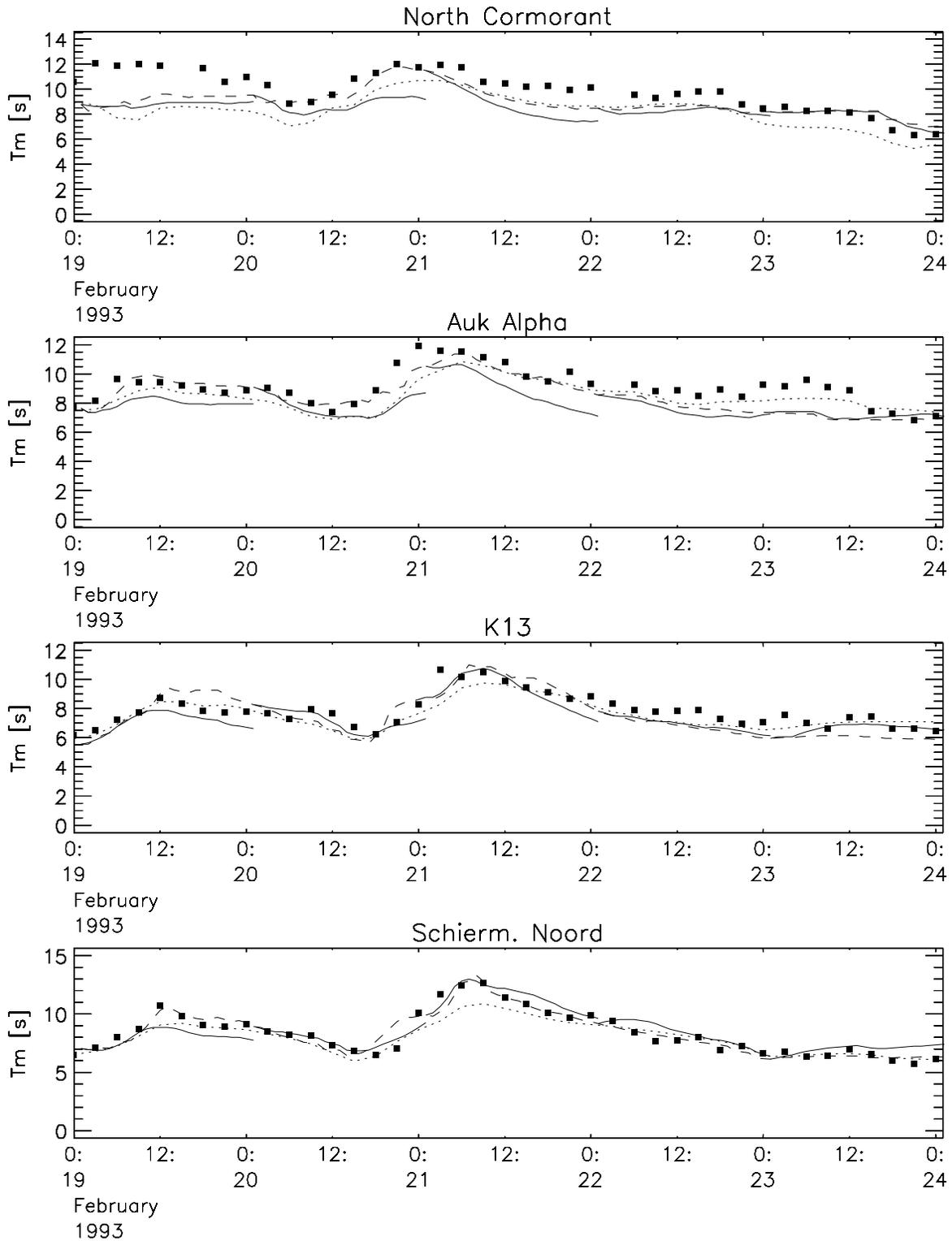,height=25cm}}
\vspace{-4cm}
\caption{ \label{fig:daswam_noda_3}
Modeled and observed mean wave period at four locations for the period
February 19-24, 1993. Line types and markers as in figure
\ref{fig:daswam_noda_2}.}
\end{figure}

Figures \ref{fig:daswam_noda_2} and \ref{fig:daswam_noda_3} show time series
of, respectively, significant wave height and mean wave period at the same
positions. Three types of model runs, all without assimilation, are
presented: the WAM NOASS analysis run, the DASWAM NOASS analysis run with
the second generation model, and the DASWAM NOASS 24 hour forecast runs with
the third generation model PHIDIAS, starting every day at 00 GMT from the
NOASS analyzed wave field. Although all model runs have been performed with
the same wind field, the results are quite different. The WAM run severely
underestimates the wave height at the peak of the storm, at all platforms.
The DASWAM NOASS analysis run, on the other hand, gives good results for
NOC, K13 and SON, and less underprediction at AUK. The DASWAM NOASS forecast
run starting from February 20, 00 GMT, gives again much lower wave heights
at the peak of the storm, in fact comparable to the heights obtained with
the WAM model. The mean period results (fig. \ref{fig:daswam_noda_3}) give
essentially the same result: good performance by the 2nd generation model,
and underestimation by the 3d generation models. The underestimation is
largest for the DASWAM NOASS forecast run. 

The underestimation of the wave height and period by the third-generation
models is at least partly explained by the underprediction of wind speed as
was found in paragraph \ref{windfield}. The absence of bias in the second-
generation runs is surprising, given the poor model wind fields. Since the
second-generation model is meant to be an approximation to PHIDIAS, it seems
that some retuning is needed. Due to the mismatch between the two models,
the assimilation of wave observations in the second-generation model will
lead to wind speed corrections which are not optimal for PHIDIAS.

\subsubsection{Assimilation runs}

\begin{figure}[t]
\centerline{\psfig{figure=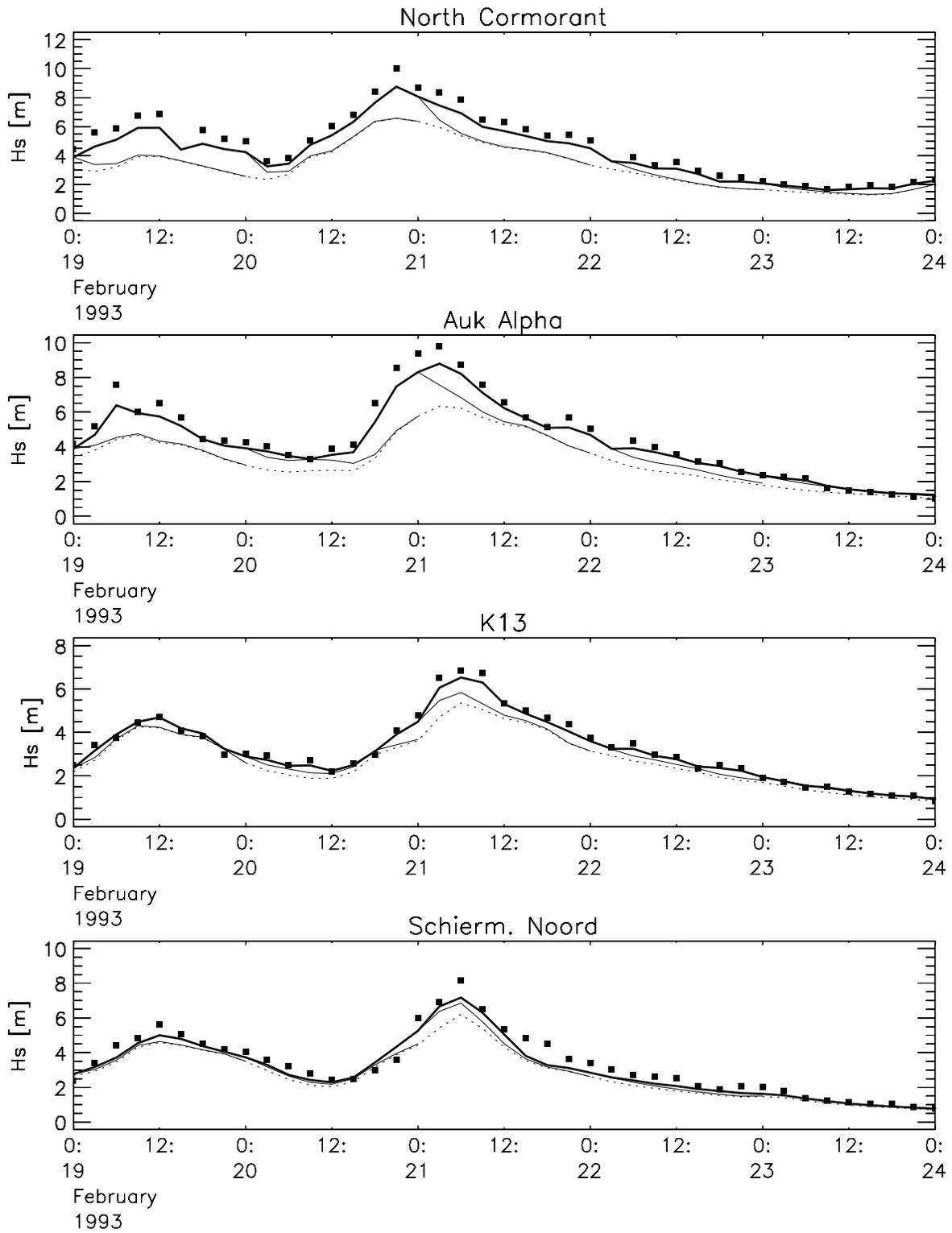,height=25cm}}
\vspace{-4cm}
\caption{ \label{fig:daswam_nosar_1}
WAM/OIP significant wave height  results for the period February 19-24,
1993. Markers: observations. Thick solid line: WAM NOSAR analysis run. Thin
solid lines: WAM NOSAR forecast runs. Dotted line: WAM NOASS run.}
\end{figure}

\begin{figure}[t]
\centerline{\psfig{figure=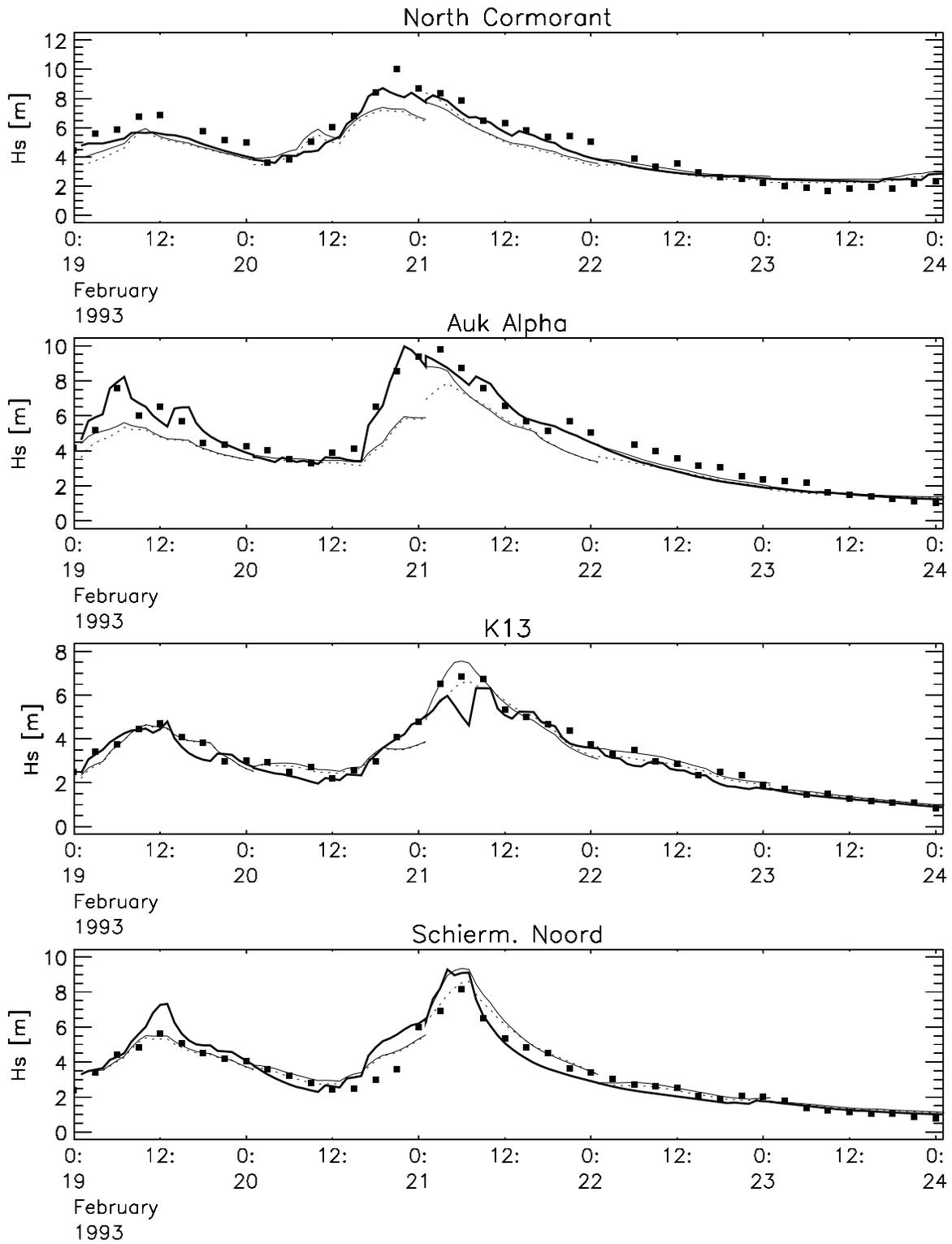,height=25cm}}
\vspace{-4cm}
\caption{ \label{fig:daswam_nosar_2}
DASWAM significant wave height results for the period February 19-24, 1993.
Markers: observations. Thick solid line: DASWAM NOSAR analysis run. Thin
solid lines: DASWAM NOSAR forecast runs. Dotted lines: DASWAM NOASS forecast
runs.}
\end{figure}

Figures \ref{fig:daswam_nosar_1} and \ref{fig:daswam_nosar_2} show NOSAR and
NOASS results, respectively for WAM/OIP and for DASWAM, versus buoy
observations. The model analyses are much closer to the observations than
the NOASS results, not only at the three locations of which the data were
assimilated, but also at SON (especially for WAM/OIP, figure
\ref{fig:daswam_nosar_1}). For both systems, the memory of the assimilation
(i.e., the time within which the NOSAR forecast relaxes back the NOASS
forecast) varies between 6 hours for the northerly stations and 12 hours for
the southerly locations, where the impact of the assimilation "upstream",
at AUK and NOC, is felt in the forecast. This period is relatively short:
for both systems, results have been reported where the impact lasted over
24 hours (Delft Hydraulics, 1995; Voorrips et al, 1996). Since wind sea
dominates in this period, improvements in the wave field are more quickly
lost than in a pure swell situation. 

\begin{figure}
\vspace{-2cm}
\centerline{\psfig{figure=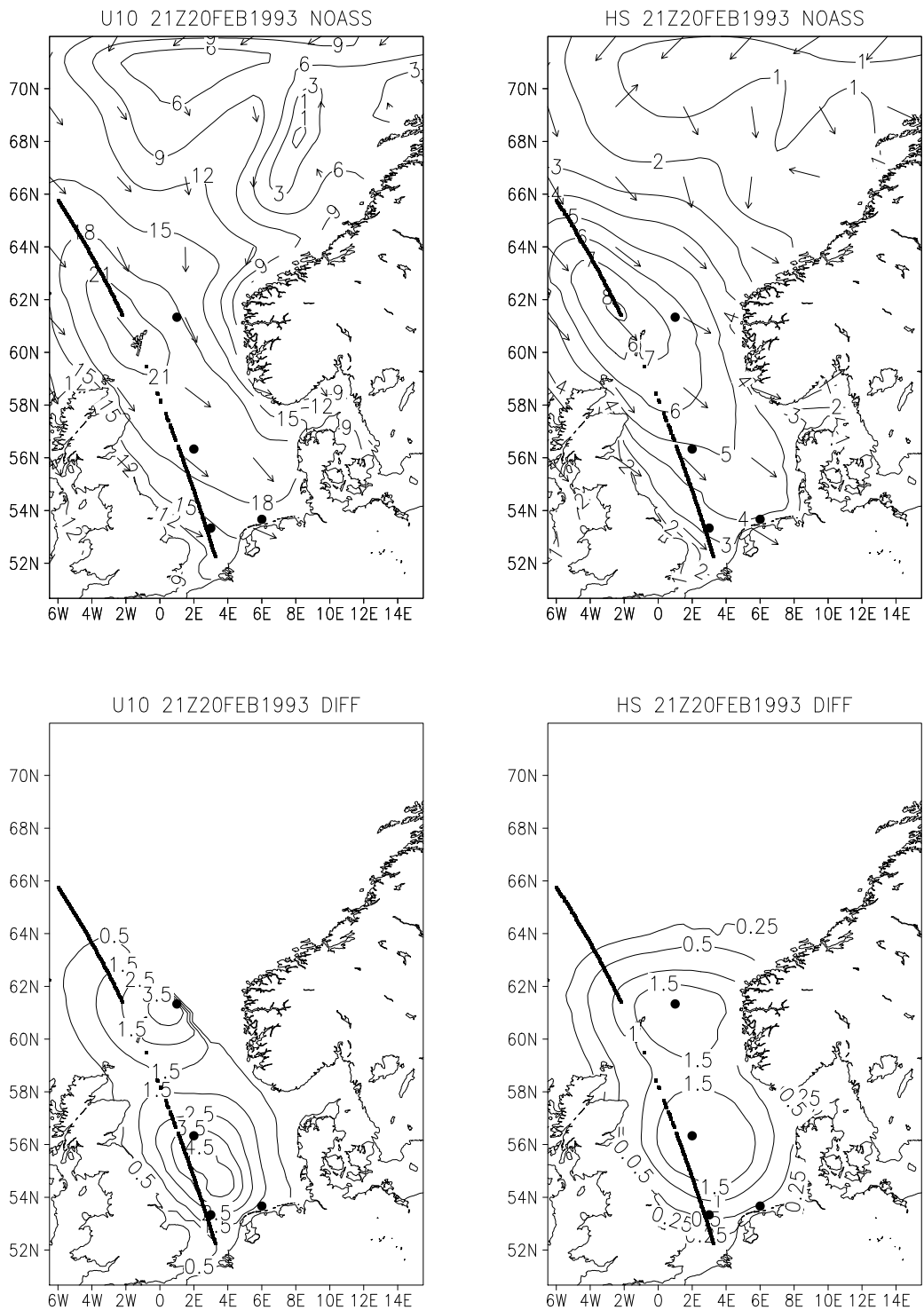,height=20cm}}
\caption{ \label{fig:daswam_map_1}
WAM/OIP model wind and wave height fields at February 21, 21 GMT. Upper
left: HIRLAM wind field. Upper right: $H_s$ for the NOASS run. Lower left:
increments of $U_{10}$ for the NOSAR assimilation run. Lower right:
increments of $H_s$ for the NOSAR run. Also indicated is the ERS-1 altimeter
track and the location of the four buoys.}
\end{figure}

\begin{figure}
\vspace{-2cm}
\centerline{\psfig{figure=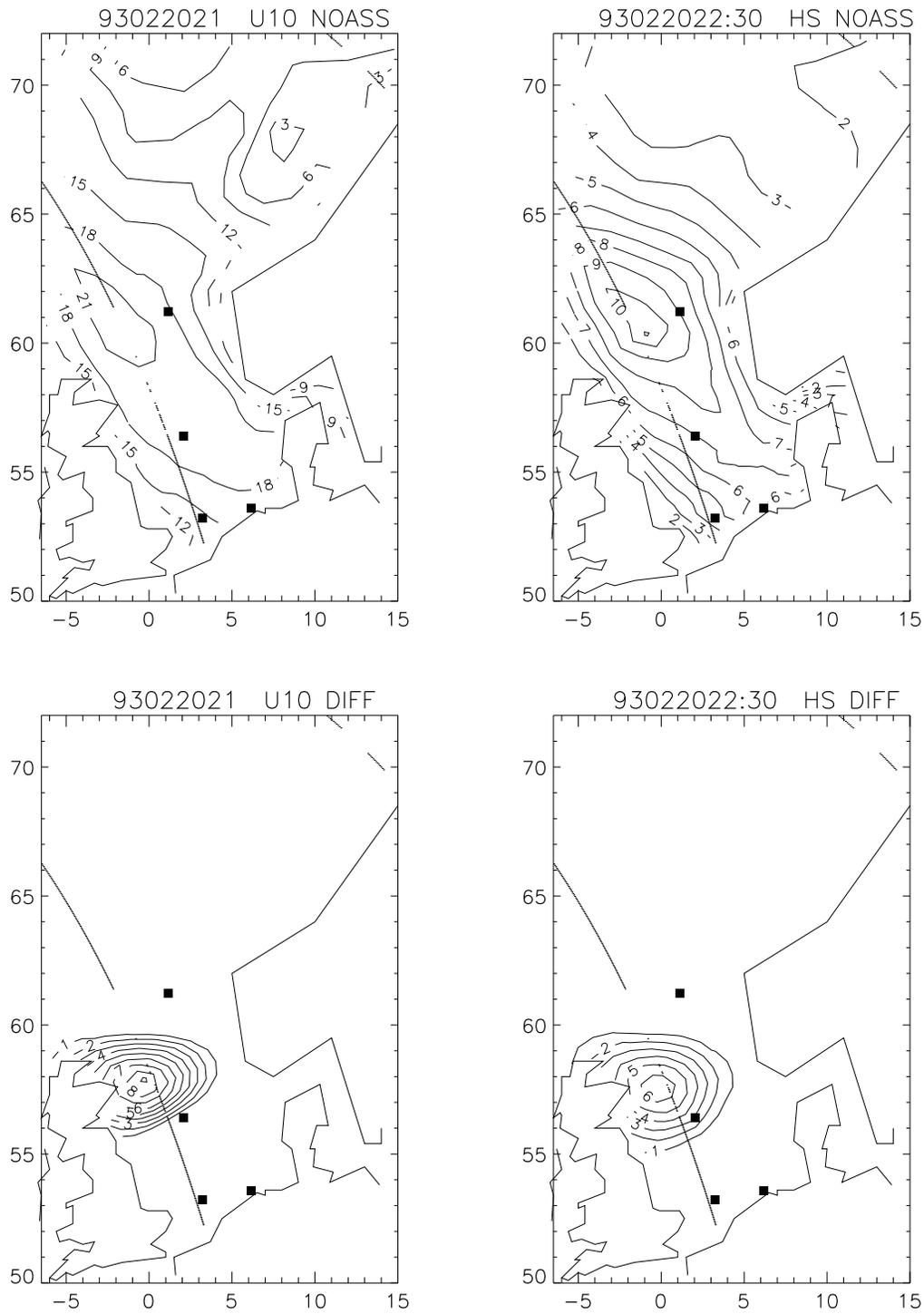,height=20cm}}
\caption{ \label{fig:daswam_map_2}
DASWAM model wind and wave height fields at February 21, 21 GMT. Upper left:
HIRLAM wind field. Upper right: $Hs$ for the NOASS run. Lower left:
increments of $U_{10}$ for the NOSAR assimilation run. Lower right:
increments of $H_s$ for the NOSAR run.}
\end{figure}

Although the analyses seem comparable when looking only at the time series
of the analyzed buoys, the nature of the assimilation is quite different.
Figure \ref{fig:daswam_map_1} shows the WAM NOASS analyzed wind and wave
fields at February 20, 21 GMT, and the increments due to the NOSAR
assimilation. The wave height increments are clearly centered around the
buoy locations, as is dictated by the fixed error covariance structure in
the OI scheme. The wind speed corrections are also quite local. North of
NOC, the wind speed is not updated, since the wave field here was computed
to be swell in this part of the model. Figure \ref{fig:daswam_map_2} shows
the same fields for DASWAM. The corrections here are entirely different. As
can be seen from figure \ref{fig:daswam_noda_2}, the DASWAM NOASS analysis
correctly predicts the wave height at NOC, but underpredicts at AUK. The
variational DASWAM system responds to this by changing its control, the wind
field, between the two stations. In this way, the model manages to increase
the wave height at AUK. Thus, although the obtained wave height at AUK is
approximately the same for the two schemes, the spatial distribution of the
update is entirely different.

\begin{figure}[t]
\centerline{\psfig{figure=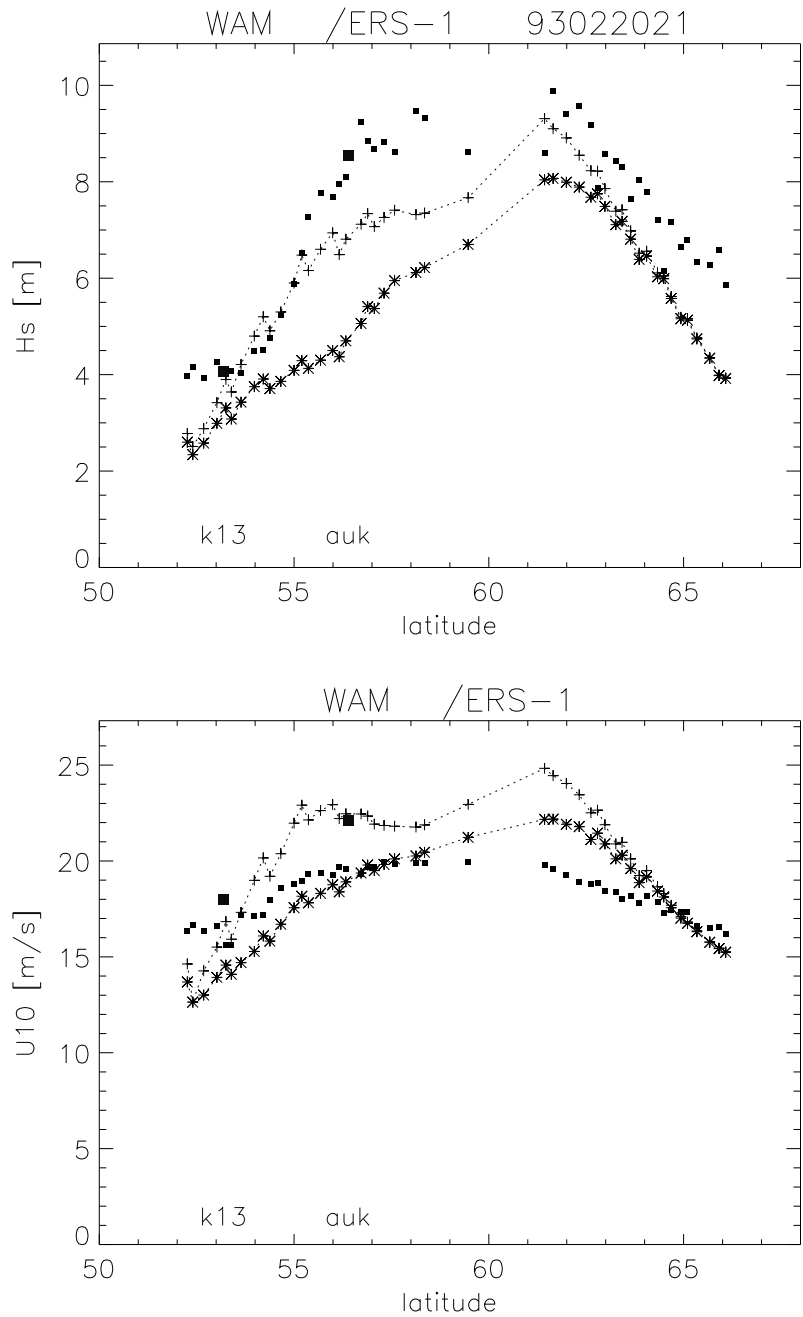,height=18cm}}
\caption{ \label{fig:daswam_alt_1}
Comparison of WAM/OIP and ERS-1 altimeter for the track of February 20, 21
GMT (the track is shown in figure \ref{fig:daswam_map_1}). Upper panel:
significant wave height. Lower panel: Wind speed. Small square markers:
altimeter observations. Stars: corresponding NOASS (analysis) model results.
Plusses: NOSAR (analysis) results. Large squares: buoy and platform
observations at K13 and AUK.}
\end{figure}

\begin{figure}[t]
\centerline{\psfig{figure=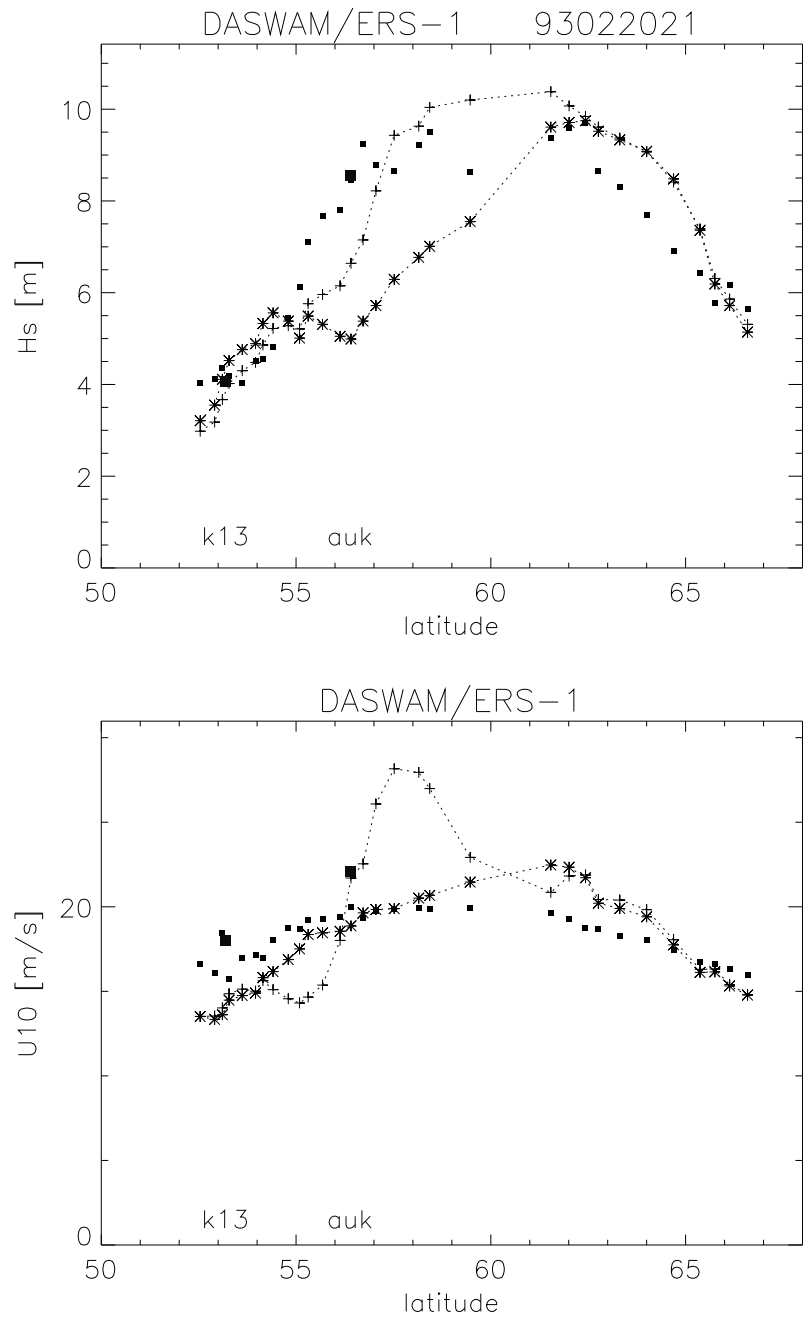,height=18cm}}
\caption{ \label{fig:daswam_alt_2}
Comparison of DASWAM and ERS-1 altimeter for the track of February 20, 21
GMT (the track is shown in figure \ref{fig:daswam_map_2}). Symbols: as in
figure \ref{fig:daswam_alt_1}.}
\end{figure}

At the time of the presented updates, the ERS-1 satellite passed. In figures
\ref{fig:daswam_map_1} and \ref{fig:daswam_map_2}, the altimeter track is
indicated, which passes close to AUK and K13. Figures \ref{fig:daswam_alt_1}
and \ref{fig:daswam_alt_2} compare the measured and modeled wind speed and
wave height along the track. Both methods manage to draw the modeled wave
height closer to the measurements after assimilation, even though the
altimeter measurements are not assimilated themselves. The wind speed is
also updated, but both methods clearly do not draw the model wind field
closer to the altimeter observations. Apparently, the assumptions underlying
the wind update are not perfect. This is even the case for DASWAM, where the
assimilation is consistent with the (second-generation) wave model, but not
with the dynamics of the atmosphere (see section \ref{daswam}).

\begin{figure}
\vspace{-2cm}
\centerline{\psfig{figure=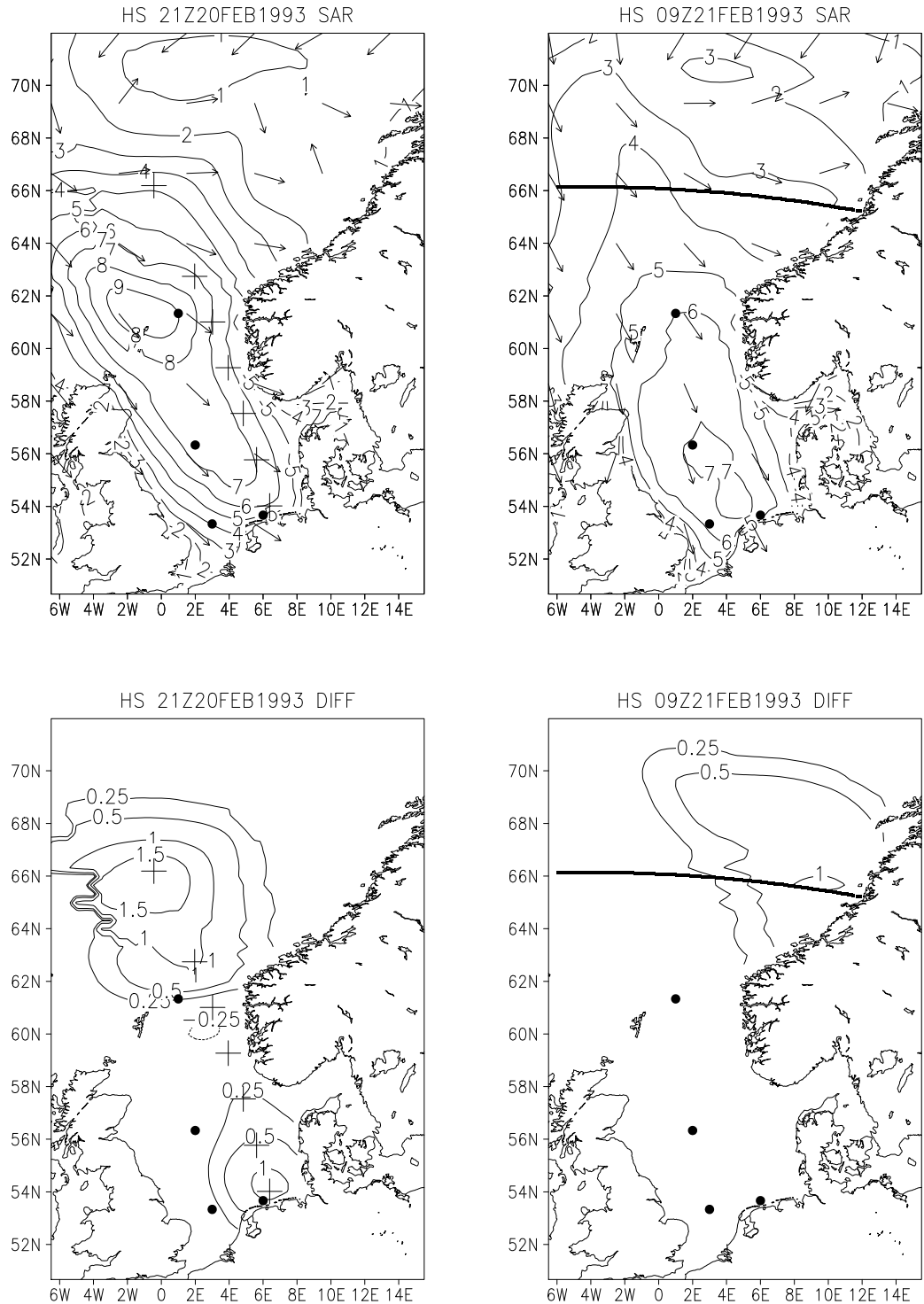,height=20cm}}
\caption{ \label{fig:daswam_map_sar}
Impact of SAR assimilation on the significant wave height with the WAM/OIP
system. Left panels: wave height fields at February 20, 21 h GMT, when the
ERS-1 passed over. Plusses indicate the SAR observations. Upper left: wave
field of the SAR analysis run. Lower left: difference between the SAR and
the NOSAR analyses at the same time. Right panels: wave fields at February
21, 9 h GMT, when the Topex-Poseidon passed. The thick sold line indicates
its altimeter track. Upper right: SAR analysis. Lower right: difference with
the NOSAR analysis.}
\end{figure}

\begin{figure}[t]
\centerline{\psfig{figure=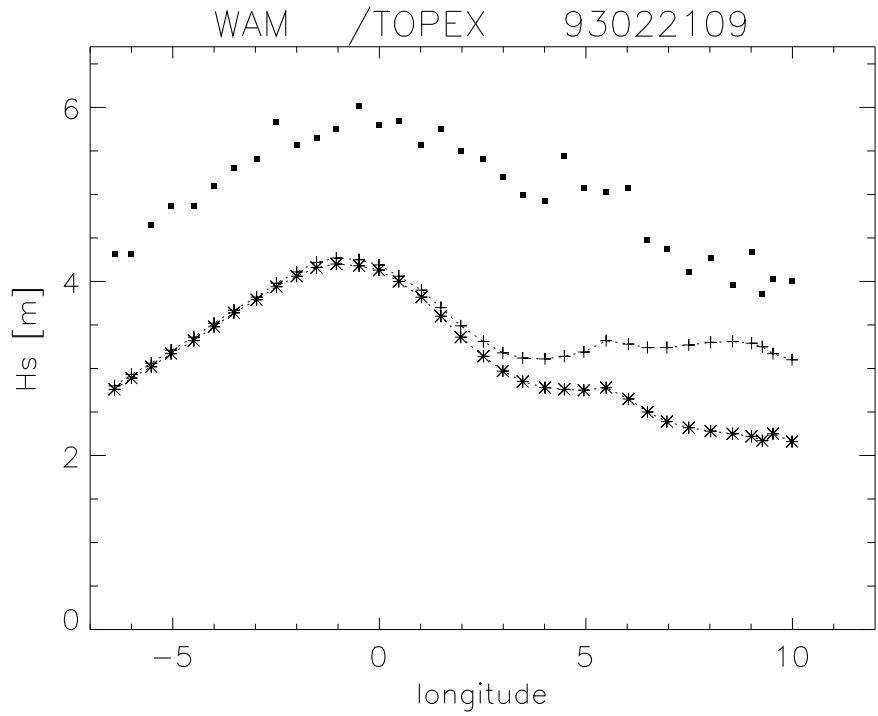,height=10cm}}
\caption{ \label{fig:daswam_sar_1}
Comparison of WAM/OIP SAR and NOSAR runs with TOPEX-POSEIDON altimeter wave
height at February 21, 9 GMT. The altimeter track is shown in figure
\ref{fig:daswam_map_sar}, right panels. Markers: altimeter wave height.
Plusses: SAR assimilation results. Stars: NOSAR assimilation results.}
\end{figure}

The impact of assimilation of the ERS-1 SAR observations on the results
during the Wadden storm was small in general. The chance that the SAR can
have an impact is not large: one needs high waves coming from the Norwegian
Sea and bad swell prediction by the model, at exactly the time that SAR
observations are made, which is only once or twice per day. In one case,
however, observations of an important track caused a clear improvement in
the Norwegian Sea. Figure \ref{fig:daswam_map_sar} shows the track at
February 20, 21 GMT, in the left panels. One can see the impact which the
SAR observations have on the WAM/OIP analyzed wave height in the Norwegian
Sea. The mean wave direction is eastward in this area, so the area of impact
travels towards the Norwegian coast. Twelve hours later, a Topex-Poseidon
track measures the wave height across the Norwegian Sea (fig.
\ref{fig:daswam_sar_1}). One can clearly see the improvement in the modeled
wave height near the Norwegian coast, because of the SAR assimilation
12~hours earlier. 

\subsection{Statistical results over the period February 19 - March 30,
1993}
\label{statistics}

\subsubsection{Buoy measurements}

The results of the analysis/forecast runs over the 39-day period February
19 - March 30 have been validated against measurements at the locations NOC,
AUK, K13 and SON. Tables \ref{tab:WAM_RMS_NOC_2} -
\ref{tab:DASWAM_RMS_SON_2} compare the NOASS (no assimilation) with the
NOSAR (assimilation of only buoy measurements) runs, in terms of the root
mean square error (RMSE)
\begin{equation}
RMSE(x) = 
\sqrt{
  \sum_{i} ( x^{mod}_i - x^{obs}_i )^2
}
\end{equation}
Five parameters are compared: significant wave height ($H_s$), low-frequency
significant wave height ($H_{10}$), mean wave period ($T_m$), mean wave
direction ($\theta_w$), and wind speed $U_{10}$. Results are given for the
last analysis time (+0 h), and for the forecast times +6 h and +12 h. In
order to reduce noise due to the relatively low number of runs (39), for
each time the results have been combined with those of 3 hours earlier,
leading to a maximum of 78 model/observed values for each forecast time. At
analysis time, the absolute value of the RMSE is given for the NOASS run.
Subsequently, for all three times the ratio of the NOSAR RMSE and the NOASS
RMSE is given, expressing the relative impact of the assimilation on the
analysis and forecast. For wind speed, only the reduction is given at
analysis time, since the assimilation schemes do not make corrections to the
forecast wind speed.

For most locations and wave parameters, the performance of the WAM and
DASWAM NOASS runs is comparable. The main difference is the larger bias in
wave height and period for WAM at the most northerly location, NOC. The
difference in RMSE here is mainly caused by the difference in bias (not
shown), which has already been noted in the discussion of the Wadden Storm
results. The bias of the HIRLAM wind speed compared to platform observations
is around -4.5 m/s at NOC, and around -1 m/s for the other locations. Again,
the underestimation of wave height and period by the WAM model seems to be
consistent with the underestimation of the wind speed. Another reason for
the larger underestimation by WAM at NOC may be that the WAM model domain
extends not as far westward as the DASWAM domain (fig.
\ref{fig:modelregions}). Consequently, external swell from the Atlantic
Ocean is more easily missed.

The impact of the assimilation on the analysis is locally much larger for
the WAM/OIP scheme than for DASWAM: for instance, the RMS error in wave
height at K13 is reduced to 37~\% of the NOASS RMSE (table
\ref{tab:WAM_RMS_K13_2}), whereas for DASWAM the reduction is only to 87~\%
(table \ref{tab:DASWAM_RMS_K13_2}). This is a direct consequence of the
different nature of the schemes: optimal interpolation draws the model to
the measurements mainly in the vicinity of the observation, whilst the
impact of the adjoint method is more global. Comparison of the impacts on
wave height at SON (not assimilated), illustrates this. The reduction in
RMSE by the WAM/OIP scheme is here 76~\% of the NOASS RMSE (table
\ref{tab:WAM_RMS_SON_2}), which is a much less dramatic improvement then at
the assimilated location K13; the reduction by DASWAM is 92~\% (table
\ref{tab:DASWAM_RMS_SON_2}), which is more comparable to the 87~\% at K13.

Also in the forecast, the impact of the WAM/OIP scheme on the RMSE is
generally higher than the impact of DASWAM, although the difference is much
smaller than at analysis time. Averaged over all four wave parameters and
over the three locations with roughly the same NOASS quality (AUK, K13,
SON), the reduction factor for the RMSE is 75~\% for WAM/OIP vs. 93~\% for
DASWAM for the +3/+6 h forecasts, and 90~\% vs 95~\% for the +9/+12 h
forecasts. Clearly, for the short term forecasts, the strong local
correction of the OIP scheme is still of importance.

In the period considered, the impact of the assimilation on the forecast is
seen up to around 12 hours. This is a shorter period than reported both for
DASWAM (Delft Hydraulics, 1995) and for WAM/OIP (Voorrips et al, 1996).
Probably, the impact is relatively small because in this period, wind sea
was in general the dominating wave system in the North Sea. In periods in
which swell is of more importance (this happens for instance in summer
periods, see Delft Hydraulics, 1995), the impact period is longer.

Both schemes also correct the wind speed during the assimilation (tables
\ref{tab:WAM_RMS_NOC_2} - \ref{tab:DASWAM_RMS_SON_2}). The correction does
not lead to a significant improvement or deterioration of the model wind
compared to the platform observations. At AUK, the reduction (WAM/OIP) or
growth (DASWAM) of the RMSE is large, but based on an insufficient number
of observations.

The impact of SAR assimilation in addition to the assimilation of buoy
observations turned out to be negligible in the comparison with the buoy
observations. This is to be expected, since especially in the neighbourhood
of the buoys, the influence of the buoy measurements is much larger than the
SAR observations, which are sparse and often far away from the buoy
location.

\subsubsection{Satellite measurements}

Tables \ref{tab:WAM_ALT} and \ref{tab:DASWAM_ALT} show a comparison of
WAM/OIP and DASWAM analyzed wave height and wind speed with measurements
from the ERS-1 and Topex-Poseidon altimeters. For the comparison two regions
are defined (fig. \ref{fig:modelregions}): region I covers the North Sea,
and region II covers part of the Norwegian Sea. All altimeter measurements
within one model grid box are averaged into one "super-observation". Since
the grids of the two model differ, the number of super-observations differs
too.

Comparison with both satellites shows a negative bias of the HIRLAM (NOASS)
wind of around 0.6~m/s in area~II, and a small positive bias (around~0.2
m/s) in region~I. These biases are smaller than those found in the platform
comparison (-4.5~m/s for NOC in area II, and around -1~m/s for AUK, K13 and
SON in area I). Differences between the TOPEX and ERS-1 results are much
smaller than those between the comparison with platform measurements (tables
\ref{tab:WAM_RMS_NOC_2}-\ref{tab:DASWAM_RMS_SON_2}) and with altimeter
measurements. This is suprising, since the calibration of the TOPEX winds
\refb{topexwinds} has been obtained by a comparison with measurements from
the same platforms (Komen et al, 1996). One way to explain the apparent
discrepancy is the assumption that the model wind speed error is not
homogeneous over the analysis areas in this relatively short period. Another
reason may be that the altimeters (TOPEX with the correction
\refb{topexwinds}) underestimate the wind speed at very high winds, like in
the Wadden Storm. For the wind speed range above 20~m/s, not enough data are
available to make a reliable validation of the altimeter wind speed
algorithms. 

For both models, the NOASS wave height in the northern region (region II)
is negatively biased compared to the altimeter measurements. This is
consistent with the buoy comparison at NOC. The negative bias is worse for
the WAM model than for DASWAM's second generation model, which is also in
agreement with the buoy comparison results. Apart from the wind speed bias,
the wave height bias may be caused by the absence of external boundary
information, which is of main importance for the Norwegian Sea (swell
entering from the Atlantic).

In the North Sea region, assimilation reduces the mean error in wave height
for both models (although the Topex-Poseidon comparison for DASWAM is not
convincing). Again, the impact caused by the WAM/OIP scheme is somewhat
larger. In the Norwegian Sea, there is still some impact for the WAM/OIP
model, but it is much smaller than in region I. The reason that the impact
is larger than for DASWAM in region II is probably the fact that the WAM
NOASS model results are more negatively biased: the impact in this region
is only the reduction of the bias.

The impact of the SAR observations is again very small. Only for WAM/OIP,
and only in region II, a small positive impact can be noted. Again, this is
the most likely candidate: region II, because the impact from buoys is
smaller for the Norwegian Sea, and WAM/OIP, because it has the largest bias
to correct.

The correction of the wind speed by the assimilation schemes has no
significant impact on the quality, when compared to observations. This
result confirms the results of the buoy comparison. 

A comparison with the ERS-1 scatterometer wind velocity data shows
essentially the same results as the altimeter wind speed measurements. 

\subsection{Discussion}
\label{discussion}
Summarizing the buoy and satellite comparisons described above, the
following main results are obtained. First, both the OIP/WAM and the DASWAM
assimilation scheme manage to improve the wave analysis field, and the short
term forecast, if not too far from the assimilation sites. Second, the OIP
scheme has a somewhat larger effect than DASWAM, both on the analysis and
on the forecast. Third, the wind speed corrections applied by both schemes
do not significantly affect the quality of the wind field. 

The second result is surprising, since the multi-time-level variational
method of DASWAM is a more sophisticated assimilation scheme than the rather
ad hoc optimal interpolation scheme OIP. The constraint of the model
evolution equations to the analysis in DASWAM should lead to a physically
consistent correction to the model parameters, while the OIP scheme does not
take the assimilation history into account. 

The fact that the wind speed corrections of DASWAM do not improve on the
(rather poor) first-guess wind field gives an indication why DASWAM does not
perform better: apparently, the choice of the control variables (wind speed)
is not optimal, or the error statistics attributed to them in the cost
function are not correct. Probably, both aspects play a role. The absence
of bias in the second-generation NOASS results given the biased wind speed,
indicates that this wave model cannot be assumed to be perfect, as is
implicated by the adjoint method (the model equations are a strong
constraint in the minimization of the cost function). Hence, the control
variables should include wave model errors as well as wind speed. Also, a
wind field representation by splines, without correlation between the
velocity components and without temporal correlation between consecutive
fields, is not optimal. 

It is not in itself a problem that the second-generation wave model used for
analysis differs from the forecast model in the DASWAM suite, because the
wind speed corrections which are determined by the assimilation are never
used by the forecast model: only the analyzed wave field after the
assimilation is used as an initial condition for PHIDIAS, the third-
generation forecast model. However, naturally the model analysis and hence
the following forecast will improve when the second-generation model is
improved. The results shown here suggest that some retuning of the model may
be useful.

The principle of variational assimilation is quite powerful, and it seems
that relatively small adaptations to the DASWAM scheme (better
representation of the error statistics, retuning of the second-generation
wave model) may greatly enhance its performance.

The OIP scheme behaved reasonably well for a rather simple assimilation
method, and from the present study it is not evident how its performance
could be improved without major modifications. Like with DASWAM, the wind
field corrections were not successful. However, in the OIP setup the wind
speed correction is only a postprocessing based on simple assumptions
(growth curves), and its significance for the wave field analysis is
limited. Neither is the wind update in comparable OI schemes necessarily
unsuccessful: in the global WAM model, some improvement of the ECMWF first-
guess winds was obtained by assimilation of ERS altimeter wave heights using
an optimal interpolation scheme (P.A.E.M. Janssen, personal communication).
The reason why the wind speed update in the present setup is not successful
may be that in these experiments, continually wave observations are
assimilated at fixed locations with very little time delay. Thus, the
assumption underlying the update scheme that at every assimilation timestep
the wave field error is in equilibrium with the wind field error, is in fact
violated.

\section{Conclusion}
\label{conclusion}

In this paper, a close comparison has been carried out between the two wave
forecast/assimilation models WAM/OIP and DASWAM, with the emphasis on the
assimilation. WAM/OIP is an example of a rather simple, optimal
interpolation method. DASWAM is a multi-time-level variational method, with
some approximations to reduce the required computation time.

The different properties of the two assimilation schemes are illustrated in
an analysis of the most interesting period which was investigated, the
"Wadden Storm" of February 20-22, 1993. The corrections of the OIP scheme
are localized around the observation sites, whereas the DASWAM scheme
corrects the wave and wind field more globally, both in space and time. The
performance of the two schemes during the Wadden storm, with respect to buoy
observations, is somewhat obscured due to the large negative bias of the
first-guess WAM run in the Norwegian Sea, and by the different behaviour of
the two wave models which play a role in the DASWAM method. The bias may be
caused either by the too low HIRLAM winds (as is supported by the platform
measurements at NOC and AUK), or by the fact that the WAM model domain does
not extend far enough to the West.

A statistical analysis of the model results against buoy measurements over
the period February 19 - March 30, 1993, shows that the OIP scheme draws the
model closer to the assimilated observations than DASWAM does. Most of the
difference disappears quickly in the forecast, but the impact of OIP remains
slightly larger at the observation locations. After 12 hours, most of the
impact of the assimilation is lost. This is probably due to the fact that
during most of the period, wind sea was the dominating wave system in the
North Sea. Comparison of the model analysis results with independent ERS-1
and Topex-Poseidon altimeter wave height measurements confirms the negative
bias of WAM in the Norwegian Sea. In the North Sea, a positive impact of the
buoy assimilation is seen in the two methods. In the Norwegian Sea, the
impact of OIP is larger, probably because the quality of the WAM first-guess
is worse.

The wind speed corrections which are the result of the assimilation, do not
improve or deteriorate significantly the quality of the wind fields, as
compared to measurements from platforms, the ERS-1 and Topex-Poseidon
altimeters, or the ERS-1 scatterometer.

The more elaborate variational scheme of DASWAM did not lead to more
accurate wave analyses or predictions. In the wind-sea dominated test
period, the better tuned statistical basis of OIP proved effective, whereas
the potential of DASWAM for making non-local adjustments did not show any
clear advantages. It is suspected that improvements in the description of
the error statistics in the cost function, together with a retuning of the
approximate wave model which is used in the analysis, may show the
advantages of the variational method more clearly.

The impact of the SAR assimilation is generally small, especially in the
North Sea, where the number of buoy observations is much larger than the
number of SAR observations. In the Norwegian Sea, some impact is shown in
the WAM/OIP scheme at analysis time. In this region, the number of
conventional observations is small, and assimilation of satellite
observations is expected to become increasingly important in the future.

\section*{Acknowledgments}

The authors would like to thank Peter Groenewoud from ARGOSS who performed
the test runs with DASWAM, Patrick Heimbach from MPIM for the supply of the
inverted ERS-1 SAR spectra, Jean Tournadre from IFREMER for the ERS-1
altimeter and scatterometer data, and Edwin Wisse from the Technical
University Delft for the Topex-Poseidon altimeter data. 

Part of this work was carried out in the framework of the European Coupled
Atmosphere/Wave/Ocean (ECAWOM) project which is funded by the MAST (Marine
Science and Technology) Programme of the European Union.


\clearpage

\begin{table} [t]
\begin{center}
\begin{tabular}{||c|c|rc||c|c|c||} \hline \hline

Parameter & $N$  
& \multicolumn{2}{c||}{NOASS RMSE}
& $\frac{NOSAR}{NOASS}$ (\%)
& $\frac{NOSAR}{NOASS}$ (\%)
& $\frac{NOSAR}{NOASS}$ (\%)
\\ 

&   
& \multicolumn{2}{c||}{-3h - 0h }
& -3h - 0h 
& +3h - +6h 
& +9h - +12h 
\\ 
\hline

$H_s        $ &     40 &    1.23 & m   &   38 &   80 &   93  \\
$H_{10}     $ &     40 &    1.31 & m   &   39 &   76 &   91  \\
$T_m        $ &     40 &    1.63 & s   &   36 &   71 &   91  \\
$\theta_{w} $ &     40 &   13.73 & deg &   59 &   78 &   93  \\
$U_{10}     $ &     40 &    5.12 & m/s &  107 & &  \\

\hline \hline
\end{tabular} 
\end{center}
\caption{ \label{tab:WAM_RMS_NOC_2}
RMS~error reduction for WAM/OIP at NOC.}
\end{table}

\begin{table} [t]
\begin{center}
\begin{tabular}{||c|c|rc||c|c|c||} \hline \hline

Parameter & $N$  
& \multicolumn{2}{c||}{NOASS RMSE}
& $\frac{NOSAR}{NOASS}$ (\%)
& $\frac{NOSAR}{NOASS}$ (\%)
& $\frac{NOSAR}{NOASS}$ (\%)
\\ 

&   
& \multicolumn{2}{c||}{-3h - 0h }
& -3h - 0h 
& +3h - +6h 
& +9h - +12h 
\\ 
\hline

$H_s        $ &     40 &    0.87 & m   &   82 &   92 &   94  \\
$H_{10}     $ &     40 &    0.85 & m   &   82 &   97 &   95  \\
$T_m        $ &     40 &    0.89 & s   &  104 &  101 &  102  \\
$\theta_{w} $ &     40 &   18.46 & deg &   80 &  101 &  102  \\
$U_{10}     $ &     40 &    5.59 & m/s &  103 & &  \\

\hline \hline
\end{tabular} 
\end{center}
\caption{ \label{tab:DASWAM_RMS_NOC_2}
RMS~error reduction for DASWAM at NOC.}
\end{table}

\begin{table} [t]
\begin{center}
\begin{tabular}{||c|c|rc||c|c|c||} \hline \hline

Parameter & $N$  
& \multicolumn{2}{c||}{NOASS RMSE}
& $\frac{NOSAR}{NOASS}$ (\%)
& $\frac{NOSAR}{NOASS}$ (\%)
& $\frac{NOSAR}{NOASS}$ (\%)
\\ 

&   
& \multicolumn{2}{c||}{-3h - 0h }
& -3h - 0h 
& +3h - +6h 
& +9h - +12h 
\\ 
\hline

$H_s        $ &     67 &    0.86 & m   &   32 &   72 &   89  \\
$H_{10}     $ &     67 &    0.91 & m   &   33 &   64 &   85  \\
$T_m        $ &     59 &    1.00 & s   &   36 &   73 &   82  \\
$\theta_{w} $ &     59 &   14.53 & deg &   51 &   96 &   99  \\
$U_{10}     $ &     13 &    2.78 & m/s &   77 & &  \\

\hline \hline
\end{tabular} 
\end{center}
\caption{ \label{tab:WAM_RMS_AUK_2}
RMS~error reduction for WAM/OIP at AUK.}
\end{table}

\begin{table} [t]
\begin{center}
\begin{tabular}{||c|c|rc||c|c|c||} \hline \hline

Parameter & $N$  
& \multicolumn{2}{c||}{NOASS RMSE}
& $\frac{NOSAR}{NOASS}$ (\%)
& $\frac{NOSAR}{NOASS}$ (\%)
& $\frac{NOSAR}{NOASS}$ (\%)
\\ 

&   
& \multicolumn{2}{c||}{-3h - 0h }
& -3h - 0h 
& +3h - +6h 
& +9h - +12h 
\\ 
\hline

$H_s        $ &     67 &    0.71 & m   &   65 &   84 &   96  \\
$H_{10}     $ &     67 &    0.87 & m   &   60 &   78 &   96  \\
$T_m        $ &     59 &    1.16 & s   &   88 &   88 &   96  \\
$\theta_{w} $ &     59 &   25.24 & deg &   86 &   95 &   92  \\
$U_{10}     $ &     13 &    2.91 & m/s &  133 & &  \\

\hline \hline
\end{tabular} 
\end{center}
\caption{ \label{tab:DASWAM_RMS_AUK_2}
RMS~error reduction for DASWAM at AUK.}
\end{table}

\clearpage

\begin{table} [t]
\begin{center}
\begin{tabular}{||c|c|rc||c|c|c||} \hline \hline

Parameter & $N$  
& \multicolumn{2}{c||}{NOASS RMSE}
& $\frac{NOSAR}{NOASS}$ (\%)
& $\frac{NOSAR}{NOASS}$ (\%)
& $\frac{NOSAR}{NOASS}$ (\%)
\\ 

&   
& \multicolumn{2}{c||}{-3h - 0h }
& -3h - 0h 
& +3h - +6h 
& +9h - +12h 
\\ 
\hline

$H_s        $ &     72 &    0.40 & m   &   37 &   69 &   91  \\
$H_{10}     $ &     72 &    0.18 & m   &   71 &   54 &   81  \\
$T_m        $ &     48 &    0.51 & s   &   47 &   63 &   83  \\
$\theta_{w} $ &     48 &   17.37 & deg &   36 &   83 &  106  \\
$U_{10}     $ &     72 &    1.92 & m/s &   98 & &  \\

\hline \hline
\end{tabular} 
\end{center}
\caption{ \label{tab:WAM_RMS_K13_2}
RMS~error reduction for WAM/OIP at K13.}
\end{table}

\begin{table} [t]
\begin{center}
\begin{tabular}{||c|c|rc||c|c|c||} \hline \hline

Parameter & $N$  
& \multicolumn{2}{c||}{NOASS RMSE}
& $\frac{NOSAR}{NOASS}$ (\%)
& $\frac{NOSAR}{NOASS}$ (\%)
& $\frac{NOSAR}{NOASS}$ (\%)
\\ 

&   
& \multicolumn{2}{c||}{-3h - 0h }
& -3h - 0h 
& +3h - +6h 
& +9h - +12h 
\\ 
\hline

$H_s        $ &     72 &    0.41 & m   &   87 &   94 &   94  \\
$H_{10}     $ &     72 &    0.33 & m   &   86 &   83 &   87  \\
$T_m        $ &     48 &    0.67 & s   &   97 &  103 &   99  \\
$\theta_{w} $ &     48 &   15.83 & deg &   96 &   97 &   98  \\
$U_{10}     $ &     72 &    2.13 & m/s &  100 & &  \\

\hline \hline
\end{tabular} 
\end{center}
\caption{ \label{tab:DASWAM_RMS_K13_2}
RMS~error reduction for DASWAM at K13.}
\end{table}

\begin{table} [t]
\begin{center}
\begin{tabular}{||c|c|rc||c|c|c||} \hline \hline

Parameter & $N$  
& \multicolumn{2}{c||}{NOASS RMSE}
& $\frac{NOSAR}{NOASS}$ (\%)
& $\frac{NOSAR}{NOASS}$ (\%)
& $\frac{NOSAR}{NOASS}$ (\%)
\\ 

&   
& \multicolumn{2}{c||}{-3h - 0h }
& -3h - 0h 
& +3h - +6h 
& +9h - +12h 
\\ 
\hline

$H_s        $ &     75 &    0.41 & m   &   76 &   76 &   89  \\
$H_{10}     $ &     75 &    0.35 & m   &   67 &   58 &   76  \\
$T_m        $ &     42 &    0.59 & s   &   94 &   68 &   89  \\
$\theta_{w} $ &     42 &   15.10 & deg &   93 &  122 &  105  \\
$U_{10}     $ &     75 &    2.37 & m/s &   98 & &  \\

\hline \hline
\end{tabular} 
\end{center}
\caption{ \label{tab:WAM_RMS_SON_2}
RMS~error reduction for WAM/OIP at SON.}
\end{table}

\begin{table} [t]
\begin{center}
\begin{tabular}{||c|c|rc||c|c|c||} \hline \hline

Parameter & $N$  
& \multicolumn{2}{c||}{NOASS RMSE}
& $\frac{NOSAR}{NOASS}$ (\%)
& $\frac{NOSAR}{NOASS}$ (\%)
& $\frac{NOSAR}{NOASS}$ (\%)
\\ 

&   
& \multicolumn{2}{c||}{-3h - 0h }
& -3h - 0h 
& +3h - +6h 
& +9h - +12h 
\\ 
\hline

$H_s        $ &     75 &    0.54 & m   &   92 &  104 &   94  \\
$H_{10}     $ &     75 &    0.47 & m   &   84 &   98 &   91  \\
$T_m        $ &     42 &    0.90 & s   &   78 &   92 &   87  \\
$\theta_{w} $ &     42 &   16.53 & deg &  104 &   97 &  104  \\
$U_{10}     $ &     75 &    2.35 & m/s &  100 & &  \\

\hline \hline
\end{tabular} 
\end{center}
\caption{ \label{tab:DASWAM_RMS_SON_2}
RMS~error reduction for DASWAM at SON.}
\end{table}

\clearpage

\begin{table} [t]
\begin{center}
\begin{tabular}{||c|c|c|r||r|r|r|r||} \hline \hline

area & satellite & run & $N$  
& \multicolumn{2}{c|}{wave height}
& \multicolumn{2}{c||}{wind speed}
\\ 
\cline{5-8}

&    &   &
& bias & $\sigma$
& bias & $\sigma$
\\ 
\hline

 I   & ERS-1 & NOASS &   340 &    -0.36 &     0.72 &     0.15 &     1.84 \\
 I   & ERS-1 & NOSAR &   340 &    -0.17 &     0.53 &     0.31 &     1.90 \\
 I   & ERS-1 & SAR   &   340 &    -0.16 &     0.53 &     0.30 &     1.89 \\
 I   & TOPEX & NOASS &   510 &    -0.26 &     0.53 &     0.47 &     1.68 \\ 
 I   & TOPEX & NOSAR &   510 &    -0.12 &     0.43 &     0.49 &     1.69 \\ 
 I   & TOPEX & SAR   &   510 &    -0.14 &     0.42 &     0.50 &     1.68 \\
\hline
 II  & ERS-1 & NOASS &   662 &    -1.37 &     0.78 &    -0.86 &     2.29 \\
 II  & ERS-1 & NOSAR &   662 &    -1.24 &     0.77 &    -0.86 &     2.28 \\
 II  & ERS-1 & SAR   &   662 &    -1.18 &     0.77 &    -0.82 &     2.44 \\
 II  & TOPEX & NOASS &  1764 &    -1.25 &     0.76 &    -0.45 &     2.37 \\
 II  & TOPEX & NOSAR &  1764 &    -1.14 &     0.75 &    -0.41 &     2.38 \\
 II  & TOPEX & SAR   &  1764 &    -1.08 &     0.73 &    -0.42 &     2.37 \\
\hline \hline
\end{tabular} 
\end{center}
\caption{ \label{tab:WAM_ALT}
Altimeter statistics for WAM/OIP.}
\end{table}

\begin{table} [t]
\begin{center}
\begin{tabular}{||c|c|c|r||r|r|r|r||} \hline \hline

area & satellite & run & $N$  
& \multicolumn{2}{c|}{wave height}
& \multicolumn{2}{c||}{wind speed}
\\ 
\cline{5-8}

&    &   &
& bias & $\sigma$
& bias & $\sigma$
\\ 
\hline

 I   & ERS-1 & NOASS &   298 &     0.02 &     0.72 &     0.22 &     1.86 \\
 I   & ERS-1 & NOSAR &   298 &     0.08 &     0.55 &     0.21 &     2.03 \\
 I   & ERS-1 & SAR   &   298 &     0.11 &     0.56 &     0.27 &     2.06 \\
 I   & TOPEX & NOASS &   496 &     0.21 &     0.64 &     0.40 &     1.67 \\
 I   & TOPEX & NOSAR &   496 &     0.18 &     0.64 &     0.34 &     1.69 \\
 I   & TOPEX & SAR   &   496 &     0.23 &     0.73 &     0.32 &     1.69 \\
\hline
 II  & ERS-1 & NOASS &   407 &    -0.53 &     0.88 &    -0.81 &     2.34 \\
 II  & ERS-1 & NOSAR &   407 &    -0.56 &     0.81 &    -1.06 &     2.32 \\
 II  & ERS-1 & SAR   &   407 &    -0.56 &     0.91 &    -1.33 &     2.28 \\
 II  & TOPEX & NOASS &   779 &    -0.30 &     0.88 &    -0.37 &     2.33 \\
 II  & TOPEX & NOSAR &   779 &    -0.38 &     0.79 &    -0.44 &     2.27 \\
 II  & TOPEX & SAR   &   779 &    -0.34 &     0.85 &    -0.40 &     2.31 \\

\hline \hline
\end{tabular} 
\end{center}
\caption{ \label{tab:DASWAM_ALT}
Altimeter statistics for DASWAM.}
\end{table}
\clearpage

\end{document}